\documentclass[11pt,a4paper]{article}
\usepackage{amsmath,amssymb,epsfig,a4,latexsym}

\usepackage{slashed}
\usepackage{jheppub}

\usepackage{epsfig,multicol,bbm}
\usepackage{axodraw2}
\newcommand\fverb{\setbox\fverbbox=\hbox\bgroup\verb}
\newcommand\fverbdo{\egroup\medskip\noindent%
            \fbox{\unhbox\fverbbox}\ }
\newcommand\fverbit{\egroup\item[\fbox{\unhbox\fverbbox}]}
\newbox\fverbbox

\newcounter{im}
\newcommand{\nn}{\nonumber}

\newcommand{\be}{\begin{equation}}
\newcommand{\ee}{\end{equation}}
\newcommand{\ba}{\begin{array}}
\newcommand{\ea}{\end{array}}
\newcommand{\bea}{\begin{eqnarray}}
\newcommand{\eea}{\end{eqnarray}}
\newcommand{\bqa}{\begin{eqnarray}}
\newcommand{\eqa}{\end{eqnarray}}

\newcommand{\bfig}{\begin{center}\begin{picture}}
\newcommand{\efig}[1]{\end{picture}\\{\small #1}\end{center}}

\usepackage{epsfig}
\usepackage{amssymb}
\usepackage{graphicx}
\usepackage{feynmp}
\usepackage{subfigure}
\usepackage{mathtools}
\usepackage{paralist}
\usepackage{slashed}
\usepackage{jheppub}
\usepackage{subfigure}
\usepackage{etoolbox}
\usepackage{array}
\usepackage{enumitem}
\usepackage{pifont}

\usepackage{slashed}
\usepackage{jheppub}
\usepackage{color}

\allowdisplaybreaks
\newcommand{\RS}{\overline{\text{MS}}}
\newcommand{\Neps}{N_\epsilon}
\newcommand{\alphas}{\alpha_s}
\newcommand{\alphae}{\alpha_e}

\newcommand{\defeq}{\mathrel{\mathop:}=}
\newcommand{\eqdef}{=\mathrel{\mathop:}}

\newcommand{\cusp}{\mbox{{\tiny cusp}}}
\newcommand{\bd}{\begin{displaymath}}
\newcommand{\ed}{\end{displaymath}}

\newcommand{\gsim}{\;\rlap{\lower 3.5 pt \hbox{$\mathchar \sim$}} \raise 1pt \hbox {$>$}\;}
\newcommand{\lsim}{\;\rlap{\lower 3.5 pt \hbox{$\mathchar \sim$}} \raise 1pt \hbox {$<$}\;}

\newcommand{\bc}{\begin{center}}
\newcommand{\ec}{\end{center}}

\def\mb{m}

\def\gsfullbare{g_{s}^{0,f}}
\def\gefullbare{g_{e}^{0,f}}
\def\gfullbare{g^{0,f}}
\def\alphaifull{\alpha_{i}^{f}}

\def\Xfullbare{X^{0,f}}
\def\Lfull{\mathcal{L}^{f}}
\def\psifullbare{\psi^{0,f}}
\def\Ahatfullbare{\Ahat^{0,f}}

\def\Atildefullbare{\Atilde^{0,f}}
\def\Pihatfullbare{\hat{\Pi}^{0,f}}
\def\Pitildefullbare{\tilde{\Pi}^{0,f}}
\def\Gammaghatfullbare{\Gamma_{\ghat q\bar{q}}^{0,f}}
\def\Gammagtildefullbare{\Gamma_{\gtilde q\bar{q}}^{0,f}}
\def\Mnfull{{\cal M}_n(\{\alpha\}^{f})}
\def\Zfull{\bar{Z}^{f}}

\def\gseffbare{g_{s}^0}
\def\geeffbare{g_{e}^0}
\def\Ahateffbare{\Ahat^0}
\def\Atildeeffbare{\Atilde^0}
\def\Pihateffbare{\hat{\Pi}^0}
\def\Pitildeeffbare{\tilde{\Pi}^0}
\def\Gammaghateffbare{\Gamma_{\ghat q\bar{q}}^{0}}
\def\Gammagtildeeffbare{\Gamma_{\gtilde q\bar{q}}^{0}}

\def\ghat{{\hat{g}}}
\def\gtilde{{\tilde{g}}}

\def\HV{{\scshape hv}}
\def\FDH{{\scshape fdh}}
\def\DRED{{\scshape dred}}
\def\CDR{{\scshape cdr}}

\def\RS{{\scshape rs}}

\def\mRS{{\rm \scriptscriptstyle RS}}

\def\Ahat{{\hat{A}}}

\def\Atilde{{\tilde{A}}}

\def\gs{g_{s}}
\def\ge{g_{e}}

\def\gammahat{\hat{\gamma}}
\def\gammatilde{\tilde{\gamma}}
\def\ghat{{\hat{g}}}
\def\gtilde{{\tilde{g}}}

\begin{document}
\begin{fmffile}{NONabFF}
\thispagestyle{empty}
\begin{flushright}
PSI-PR-16-11\\
ZU-TH 26/16\\
\today\\
\end{flushright}
\vspace{3em}
\begin{center}
{\Large\bf Regularization-scheme dependence of QCD amplitudes in the massive case}
\\
\vspace{3em}
{\sc Ch.~Gnendiger$^a$, A.~Signer$^{a,b}$, A.~Visconti$^{a,b}$
}\\[2em]
{\sl ${}^a$ Paul Scherrer Institut,\\
CH-5232 Villigen PSI, Switzerland \\
\vspace{0.3cm}
${}^b$ Physik-Institut, Universit\"at Z\"urich, \\
Winterthurerstrasse 190,
CH-8057 Z\"urich, Switzerland}
\setcounter{footnote}{0}
\end{center}
\vspace{2ex}
\begin{abstract}
{} We investigate QCD amplitudes with massive quarks computed in the
four-dimensional helicity scheme (FDH) and dimensional reduction at
NNLO and describe how they are related to the corresponding amplitudes
computed in conventional dimensional regularization. To this end, the
scheme dependence of the heavy quark and the velocity-dependent cusp
anomalous dimensions is determined using soft-collinear effective
theory. The results are checked against explicit computations of
massive form factors in FDH at NNLO. Our results complete the
description of the scheme dependence of QCD amplitudes at NNLO.
\end{abstract}

\vspace{0.5cm}
\centerline

\newpage
\setcounter{page}{1}

 \noindent\hrulefill
 \tableofcontents
 \noindent\hrulefill

\bigskip
\section{Introduction}
\label{sec:introduction}

The most common procedure to regularize ultraviolet (UV) and infrared
(IR) singularities of scattering amplitudes is to apply conventional
dimensional regularization (\CDR), whereby all relevant quantities are
treated as $D=4-2\epsilon$ dimensional. In \CDR, IR singularities of
next-to-next-to leading order (NNLO) scattering amplitudes in massless
QCD have a remarkably simple structure~\cite{Becher:2009cu, Gardi:2009qi,
Becher:2009qa, Gardi:2009zv}. Key ingredients are the
cusp anomalous dimension $\gamma_{\text{cusp}}$ and the anomalous
dimensions of quarks and gluons, $\gamma_{q}$ and $\gamma_{g}$,
respectively. 

For practical computations it is sometimes advantageous to apply
certain variants of \CDR, such as the 't~Hooft-Veltman scheme
(\HV)~\cite{'tHooft:1972fi}, dimensional reduction
(\DRED)~\cite{Siegel:1979wq} or the four-dimensional helicity scheme
(\FDH)~\cite{BernZviKosower:1992}. This leads to the question how
virtual amplitudes computed in these schemes are related to the
corresponding amplitudes computed in \CDR. In the massless case at
NNLO, this question has been answered in Ref~\cite{Broggio:2015dga},
where, drawing on earlier results~\cite{ Kunszt:1994,
  Catani:1996pk, Catani:1998bh, Stockinger:2005gx, Signer:2005, Signer:2008va,
  Kilgore:2011ta, Kilgore:2012tb, Gnendiger:2014nxa, Broggio:2015ata},
it has been shown that the IR structure of \CDR\ is only modified through
changes in the anomalous dimensions. We indicate this regularization-scheme
(\RS) dependence by the shifts
$\gamma_{\text{cusp}}\to\gamma_{\rm cusp}^{\mRS}$,
$\gamma_{q}\to\gamma_{q}^{\mRS}$ and $\gamma_{g}\to\gamma_{g}^{\mRS}$.
The explicit expressions of the anomalous dimensions as well as the
$\beta$~functions of the various couplings in the different schemes
have been determined at least up to NNLO.

In the presence of massive quarks there are additional structures in
the IR singularities of QCD amplitudes~\cite{Becher:2009kw}. Hence,
the scheme dependence will also have to be generalized.
At NLO the scheme-dependence has been discussed in
Ref.~\cite{Catani:2000ef}.
The generalization of the scheme dependence at NNLO to QCD amplitudes
including massive quarks is the main result of this paper. As we will
show, once the scheme-dependent UV renormalization has been carried
out, this scheme dependence is contained entirely in two additional
anomalous dimensions, the velocity-dependent cusp anomalous dimension
$\gamma_{\rm cusp}^{\mRS}(\beta)$ and the anomalous dimension of a
heavy quark $\gamma_Q^{\mRS}$.
In fact, the scheme dependence of
$\gamma_{\rm cusp}^{\mRS}(\beta)$ itself is induced solely through the
scheme dependence of the cusp anomalous dimension
$\gamma_{\rm cusp}^{\mRS}$ from the massless case.

With the results presented here it is possible to convert any NNLO QCD
amplitude between the four schemes \CDR, \HV, \FDH, and \DRED. This
allows for using whatever scheme is most convenient in the computation
of the virtual amplitude and then combine this with the real
corrections, typically computed in \CDR. In fact, for the
generalization to the massive case it is sufficient to consider the
difference between the \FDH\ and the \HV\ (or \CDR) scheme. If there
are no external gluons, \FDH\ is equivalent to \DRED.
Hence, the IR anomalous dimensions are the same, e.\,g.\
$\gamma^{\text{\FDH}}_{\text{cusp}}(\beta)
=\gamma^{\text{\DRED}}_{\text{cusp}}(\beta)$ and
$\gamma^{\text{\FDH}}_{Q}=\gamma^{\text{\DRED}}_{Q}$.
Furthermore, \CDR\ and \HV\ also have the same anomalous dimensions,
$\gamma^{\text{\CDR}}_{\text{cusp}}(\beta)
=\gamma^{\text{\HV}}_{\text{cusp}}(\beta)$ and
$\gamma^{\text{\CDR}}_{Q}=\gamma^{\text{\HV}}_{Q}$.
These schemes differ simply in the dimension of the polarization sum of
external gluons.

Apart from the four schemes treated in this paper, other possibilities
to regularize virtual amplitudes have been considered.  The
\FDH\ scheme has been adapted to the so-called {\scshape fdf} scheme
(four-dimensional formulation) for using unitary-based methods to
compute NLO amplitudes~\cite{Fazio:2014xea,Bobadilla:2015wma}. There
are also proposals to abandon dimensional regularization altogether
and perform computations completely in four dimensions in the context
of implicit regularization~\cite{Pontes:2006br, Dias:2008iz,
  Fargnoli:2010ec, Cherchiglia:2010yd}, {\scshape fdr}
(four-dimensional regularization/renormalization)~\cite{Pittau:2012zd,
  Donati:2013voa, Zirke:2015spg}, and using loop-tree duality to deal
with IR singularities at the integrand
level~\cite{Hernandez-Pinto:2015ysa, Sborlini:2015uia,
  Sborlini:2016fcj}.

While this list is by no means exhaustive it shows that despite the
impressive technical advances in computing higher-order corrections in
\CDR\ there is considerable interest in exploring alternative methods.
The results presented here complete the description at NNLO of a first
step away from a fully $D$ dimensional treatment of the problem.
Apart from allowing to perform computations in \FDH\ and \DRED, we
hope it also helps to understand better the relation between \CDR\ and
the different four-dimensional approaches mentioned above. The
ultimate goal is, of course, to develop efficient methods to
explicitly perform ever more accurate computations.

The paper is organized as follows: in Section~\ref{sec:schemes} we
briefly review the various schemes, the IR structure of amplitudes and
its extension to the massive case. We also discuss the UV
renormalization, emphasizing the special features of \FDH\ in the
presence of massive quarks. Section~\ref{sec:ad} is devoted to the
computation of $\gamma_Q^{\mRS}$ and $\gamma_{\rm cusp}^{\mRS}(\beta)$
at NNLO in the \FDH\ scheme. These results are obtained by direct
computations using soft-collinear effective theory.  In order to
obtain an independent test of the scheme dependence of NNLO
amplitudes, in Section~\ref{sec:examples} we compare the heavy-quark
and heavy-to-light form factors in the \FDH\ and \CDR\ schemes and
verify that the results are in agreement with the expected scheme
dependence obtained from the anomalous dimensions. We also provide a
guide on how to actually perform computations in the \FDH\ scheme and
show that the modifications compared to \CDR\ are minimal.  Finally we
present our conclusion in Section~\ref{sec:conc}.

\section{UV and IR structure of massive QCD}
\label{sec:schemes}

\subsection{DRED and FDH}

As has been shown in a series of papers \cite{Jack:1994bn,
  Stockinger:2005gx, Gnendiger:2014nxa,
  Broggio:2015ata,Broggio:2015dga}, a consistent formulation of the
dimensional reduction (\DRED) and the four-dimensional helicity (\FDH)
scheme in the framework of massless QCD requires the introduction of
three vector spaces.  In this work we investigate how this can be
extended to the case of massive partons.  In doing so we do not
consider processes including external vector fields. The names
\FDH\ and \DRED\ are in the following therefore used synonymously,
meaning that whenever a statement about the \FDH\ schemes is made, the
same argument also applies in \DRED. For a detailed discussion and a
precise definition of the schemes, of the related vector spaces and
their algebraic relations we refer to Ref.~\cite{Signer:2008va}. Here
we only provide the most important characteristics.

In \FDH, the underlying quasi $4$-dimensional space $Q4S$ with metric
$g^{\mu\nu}$ is split into a direct sum of the quasi $D$-dimensional
space of \CDR\ with metric $\ghat^{\mu\nu}$ and a disjoint space
$Q2\epsilon S$ with metric $\gtilde^{\mu\nu}$:
\begin{align}
g^{\mu\nu}=\ghat^{\mu\nu}+\gtilde^{\mu\nu} \, .
\end{align}
In order to have full control over the contributions originating from
$Q2\epsilon S$, we \textit{define} complete contractions of the corresponding
metric tensors as
\begin{align}
\gtilde^{\mu\nu}\gtilde_{\mu\nu}\defeq\Neps  \, .
\end{align}
As a consequence, arbitrary \FDH\ quantities in general depend on $\Neps$.
They are in the following denoted by a bar.

At the level of the Lagrangian, the structure of the different vector
spaces is reflected in a split of the quasi $4$-dimensional gluon
field into a $D$-dimensional gluon field and an $\epsilon$-scalar
field: $A^\mu=\Ahat^\mu+\Atilde^\mu$.  The 'particles' associated with
these fields are in the following denoted by $g$ and $\gtilde$,
respectively.  In
Refs.~\cite{Jack:1993ws,Harlander:2006rj,Kilgore:2011ta}, it has been
shown that because of this split in principle five different couplings
need to be distinguished in the bare theory: the gauge coupling
$\alphas=g_s^2/(4\pi)$, the $\gtilde q \bar{q}$ coupling
$\alphae=g_e^2/(4\pi)$, and three different quartic
$\gtilde$-couplings.  However, for the calculations presented in this
work it is sufficient to consider only $\alphas$ and $\alphae$.

For later purposes it turns out to be useful to include repeatedly
occurring universal factors in the definition of the bare couplings
\begin{equation}
a_i(m^2)\defeq
e^{-\epsilon \gamma_{E}}(4\pi)^{\epsilon}\Big{(}\frac{1}{m^{2}}\Big{)}^{\epsilon}
  \Big(\frac{\alpha_{i}^{0}}{4\pi}\Big)
=\Big{(}\frac{\mu^{2}}{m^{2}}\Big{)}^{\epsilon}
  \bar{Z}_{\alpha_{i}}\Big(\frac{\alpha_{i}}{4\pi}\Big)\, ,
\label{eq:asdef}
\end{equation}
where $\gamma_E$ is the Euler-Mascheroni constant, $m$ is the mass of
a heavy fermion, and $a_i\in\{a_s,a_e\}$.  As renormalization
prescription for the couplings we use the $\overline{\text{MS}}$
scheme throughout this work.  The constants $\bar{Z}_{\alpha_{i}}$ in
\FDH\ are given in e.\,g.\ Ref.~\cite{Gnendiger:2014nxa}.  The
perturbative expansion of \FDH/\DRED\ quantities in terms of the UV
renormalized couplings is in the following written as
\begin{align} \label{eq:generalexApp}
X^{\text{\FDH/\DRED}}(\{\alpha\},\Neps) =
\bar{X}(\{\alpha\},\Neps) \equiv \sum^{\infty}_{m,n}
\left(\frac{\alpha_s}{4 \pi}\right)^{m}
\left(\frac{\alpha_e}{4 \pi}\right)^{n}\,
\bar{X}_{mn}(\Neps) \, .
\end{align}

\subsection{IR factorization at NNLO in the FDH scheme}

In \CDR, the IR divergence structure of scattering amplitudes
including massive external partons has been investigated up to the
two-loop level in Ref.~\cite{Becher:2009kw}. Using a combination of
soft-collinear effective theory (SCET) (for an introduction see
e.\,g.\ Ref.~\cite{Becher:2014oda}) and heavy-quark effective theory
(HQET) (for an introduction see e.\,g.\ Ref.~\cite{Neubert:1993mb}) it has
been shown that amplitudes with an arbitrary number of massive and
massless legs factorize into a hard and a soft function, where the
latter depends on both massive and massless Wilson lines. For
amplitudes including massive partons, the corresponding IR anomalous
dimension has less constraints compared to the massless case and
additional color structures arise.

Starting from the \CDR\ expression for the IR anomalous dimension, we
write the two-parton correlation terms of the respective quantity in
\FDH\ as \bea \label{eq:G2parmass} \bar{\mathbf{\Gamma}}\left( \{
\underline{p} \} , \{ \underline{m} \} ,\mu \right)
\Big|_{\text{2-parton}} &=& \sum_{(i,j)} \frac{\mathbf{T}_i
  \cdot\mathbf{T}_j}{2} \bar{\gamma}_{\cusp} (\{\alpha\})
\ln{\frac{\mu^2}{-s_{ij}}} + \sum_i \bar{\gamma}_i (\{\alpha\}) \nn
\\* && - \sum_{(IJ)} \frac{\mathbf{T}_I
  \cdot\mathbf{T}_J}{2}\bar{\gamma}_{\cusp} (\beta_{IJ},\{\alpha\}) +
\sum_I \bar{\gamma}_I (\{\alpha\}) \nn \\* &&+\sum_{(Ij)}
\frac{\mathbf{T}_I \cdot\mathbf{T}_j}{2} \bar{\gamma}_{\cusp}
(\{\alpha\})\ln{\frac{m_I\, \mu}{-s_{Ij}}} \, , \eea
where the capital indices $I,J$ correspond to the massive partons and
the angle $\beta_{IJ}$ is defined as
\begin{equation}
\beta_{IJ} \defeq \text{arcosh}\left(\frac{-s_{IJ}}{2\,m_{I}m_{J}}\right) \, .
\label{eq:betaIJ}
\end{equation}
For the definition of the color generators $\mathbf{T}_{i}$, of the
kinematic variable $s_{ij}$, and of the sets $\{ \underline{p} \}$,
$\{ \underline{m} \}$ we refer to \cite{Becher:2009kw}.

In Eq.~\eqref{eq:G2parmass}, the first line corresponds to
contributions from massless partons, already discussed in
Refs.~\cite{Gnendiger:2014nxa, Broggio:2015ata,Broggio:2015dga}; the
remainder is given by additional terms arising in the massive theory.
Suppressing the dependence on the couplings, the complete set of IR
anomalous dimensions in \FDH/\DRED\ is given by
\begin{subequations}
\label{eq:IRanomDim}
\begin{align}
&\bar{\gamma}_{\cusp},\qquad\phantom{(\beta)}
\bar{\gamma}_i\in\{\bar{\gamma}_{q},\bar{\gamma}_{g},\bar{\gamma}_{\gtilde}\},
\phantom{\Big|}
\label{eq:IRanomDimMassless}
\\
&\bar{\gamma}_{\cusp}(\beta_{IJ}),\qquad
\bar{\gamma}_I\in\{\bar{\gamma}_{Q}\} \, ,
\label{eq:IRanomDimMassive}
\end{align}
\end{subequations}
where $\bar{\gamma}_{\gtilde}$ only appears in \DRED.
The quantities in the first line have been computed up to the two-loop
level in Refs.~\cite{Gnendiger:2014nxa,
  Broggio:2015ata,Broggio:2015dga}; the values of
$\bar{\gamma}_{\cusp}(\beta_{IJ})$ and $\bar{\gamma}_{Q}$
are so far unknown and will be given in Section~\ref{sec:ad}.
Since there is no difference between the IR anomalous dimensions appearing
both in \FDH\ and \DRED, relation~\eqref{eq:G2parmass} also holds in \DRED.

In the massive theory, the IR anomalous dimension also
contains three-parton correlation terms which we write in \FDH\ as
\bea
\bar{\mathbf{\Gamma}}
  \left( \{\underline{p}\}, \{\underline{m}\}, \mu \right)\Big|_{\text{3-partons}}
&=& i f^{abc} \sum_{(I,J,K)} \mathbf{T}_I^a\mathbf{T}_J^b\mathbf{T}_K^c\,
  F_1\left(\beta_{IJ}, \beta_{JK}, \beta_{KI}\right)
\nn \\*
& &
+ i f^{abc} \sum_{(I,J)}\sum_{k} \mathbf{T}_I^a\mathbf{T}_J^b\mathbf{T}_k^c\,
  f_2\left(\beta_{IJ},
    \ln\frac{-\sigma_{Ik} \, v_I \cdot p_k}{-\sigma_{Jk} \, v_J \cdot p_k}
 \right) \, ,
\label{eq:3part}
\eea
including the four-velocities of the massive partons
\begin{align}
v^\mu_I \defeq  \frac{p_I^\mu}{m_I} \, , \qquad\quad v_I^2 \equiv 1 \, .
\end{align}
In Refs.~\cite{Ferroglia:2009ep, Ferroglia:2009ii}, the functions
$F_1$ and $f_2$ are given for the case of \CDR. Since in \FDH\ these
functions do not receive evanescent contributions from the
$\epsilon$-scalar up to NNLO, Eq.~\eqref{eq:3part} is a
scheme-independent quantity at this order.  Its value in \FDH\ is
therefore the same as in \CDR.

In analogy to the massless case \cite{Gnendiger:2014nxa,
  Broggio:2015dga}, we subtract all IR divergences of QCD loop
amplitudes by means of a factor $\bar{\mathbf{Z}}$ which is given by a
path-ordered integral over $\bar{\mathbf{\Gamma}}$ (compare with
Eqs.~(2.8) and (2.12) of Ref.~\cite{Broggio:2015dga}).  This
renormalization factor is given in the effective theory where the
heavy quarks have been integrated out. Hence, it is written in terms
of $\alpha_i$, the couplings defined in the massless theory.  In the
massive case, however, we also need to take into account contributions
from heavy-quark loops. To reproduce the correct IR behavior of the
effective low-energy theory we therefore have to perform a matching of
the couplings between the full and the effective theory. For an
amplitude describing a process with $n$ external partons then the
following relation holds:
\begin{align}
 \lim_{\epsilon\to 0}\
   \bar{\mathbf{Z}}^{-1}(\{\alpha\})\,\Bigg[
   \big|\Mnfull\rangle
   \Bigg]_{\alphaifull\to\,\zeta_{\alpha_i}\alpha_i} =\
   \mbox{finite } .
   \label{eq:ZmassiveFDH}
\end{align}
As mentioned above, $\alpha_i$ is a coupling in the effective theory,
meaning that the heavy quark flavors have been integrated out. It is
related to the corresponding coupling of the full theory via the
decoupling relation $\alphaifull=\zeta_{\alpha_i}\alpha_i$. Explicit
results for the decoupling constants in the \FDH\ scheme will be given
in Section~\ref{sec:dec}.

\subsection{Mass renormalization of the $\epsilon$-scalar}
\label{sec:epsilonmass}

\begin{figure}
\begin{center}
\scalebox{.9}{
\begin{picture}(135,50)(0,25)
\DashLine(0,45)(35,45){4}
\DashLine(85,45)(120,45){4}
\DoubleArc[arrow](60,45)(25,0,180){2}
\DoubleArc[arrow](60,45)(25,180,360){2}
\Vertex(35,45){2}
\Vertex(85,45){2}
\end{picture}
}
\end{center}
\caption{\label{epsilontwopoint}
One-loop diagram that effectively generates an $\epsilon$-scalar mass
at the one-loop level. Massive quarks are depicted by double lines.}
\end{figure}
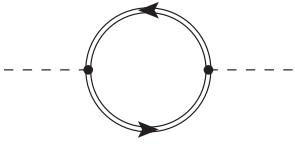

In the case of massive fermions there is no symmetry that protects the
propagator of the $\epsilon$-scalar from acquiring a mass term
$\propto m^{2}\gtilde^{\mu\nu}$ where $m$ is a fermion mass.  As a
consequence, the $\epsilon$-scalar mass is effectively shifted away
from zero, even if the $\epsilon$-scalar is massless at the
tree-level. Therefore we have to introduce a mass counterterm $\delta
m^2_{\epsilon}$ in the Lagrangian to restore the initial 'on-shell'
condition of a vanishing $\epsilon$-scalar mass~\cite{Jack:1994rk}.

At the one-loop level there is only one diagram that effectively
generates a mass term in the $\epsilon$-scalar propagator, see
Fig.~\ref{epsilontwopoint}.  To obtain the mass counterterm we need to
compute the full one-particle irreducible (1PI) two-point function of
the $\epsilon$-scalar, whose tensor structure is given by
\begin{align}
-i\tilde{\Pi}^{\mu\nu}\,
=-i\tilde{\Pi}\,p^2\,\tilde{g}^{\mu\nu}
=-i\Big(A+\frac{m^2}{p^2}B\Big)\,p^{2}\,\tilde{g}^{\mu\nu} \, ,
\end{align}
including the dimensionless quantities $A$ and $B$.  The mass
counterterm can be extracted by writing the propagator of the
$\epsilon$-scalar as
\begin{align}
\frac{-i\tilde{g}_{\mu\nu}}{p^{2}\,\big(1+\tilde{\Pi}\big)
  +\delta m^2_{\epsilon}}
=\frac{-i\tilde{g}_{\mu\nu}}{p^{2}\,\big(1+A\big)+m^2\,B+\delta m^2_{\epsilon}}\,.
\end{align}
In order to maintain the $\epsilon$-scalar massless we then require
\begin{equation}
\delta m^2_{\epsilon}
\defeq -\,m^{2}\,B
=-\,a_e(m^{2})\,m^{2}\,N_{H}\,\Bigg[
  \frac{2}{\epsilon}
  +2
  +\epsilon\Big(2+\frac{\pi^{2}}{6}\Big)
  +\mathcal{O}(\epsilon^2)
\Bigg]
+\mathcal{O}(a^2)\, ,
\label{eq:epsMassCT}
\end{equation}
where $N_{H}$ denotes the number of heavy quark flavors and the
coupling is defined in Eq.~\eqref{eq:asdef}.  As a consequence, any
time we encounter a massive loop diagram insertion as in
Fig.~\ref{epsilontwopoint}, we add the mass counterterm
\eqref{eq:epsMassCT} in order to impose the on-shell condition of a
massless $\epsilon$-scalar.

\subsection{Decoupling transformations}
\label{sec:dec}

The decoupling transformation needed in Eq.~\eqref{eq:ZmassiveFDH} is
well known for the gauge coupling. In order to extend it to $\alpha_e$
we apply the procedure described in Ref.~\cite{Chetyrkin:1997un} and
build an effective Lagrangian in which the heavy quark flavors have
been integrated out. As a consequence, the parameters and fields of
the effective theory are in general different from the ones of the
full theory.  To relate the two theories we introduce decoupling
constants in the following way:
\begin{align}
 \gfullbare=\zeta^{0}_{g}\,g^{0}, \qquad
 \Xfullbare=\sqrt{\zeta^{0}_{X}}\,X^{0} \, ,
\end{align}
where $g$ and $X$ stand for parameters and fields of the theory,
respectively.  In this way we are able to relate the full and the
effective bare QCD Lagrangian in terms of the re-scaled parameters and
fields
\begin{equation}
\Lfull\,(
  \gsfullbare,\,\gefullbare,\,\Ahatfullbare,\,\Atildefullbare,\psifullbare,
  \dots)
=\mathcal{L}\,(
  \gseffbare,\,\geeffbare,\,\Ahateffbare,\,\Atildeeffbare,\,\psi^{0},
  \dots,\{\zeta^{0}_{g}\},\{\zeta_{X}^{0}\})\, .
\end{equation}
The decoupling constants can be obtained from a matching calculation.
For $\zeta_{\Ahat}^{0}$ which is related to the gluon field decoupling,
for example, we get
\begin{subequations}
\label{eq:decgluFull}
\begin{align}
  \frac{
  -\ghat_{\mu\nu}
  }{p^2\left(1+\Pihatfullbare\right)}
  &=
  i\int {\rm d}^4x\, e^{i\, p x}\, \langle\,
    T\, \Ahatfullbare_{\mu}(x)\, \Ahatfullbare_{\nu}(0)
    \rangle
  \\*
  &=
  i\, \zeta_{\Ahat}^0 \int {\rm d}^4x\, e^{i\, p x}\,\langle\,
    T\, \Ahat^{0}_\mu(x)\, \Ahat^{0}_\nu(0)
    \rangle
  = \zeta_{\Ahat}^0\,\frac{
  -\ghat_{\mu\nu}
  }{p^2\left(1+\Pihateffbare\right)}\, ,\phantom{\Bigg|}
  \label{eq:decglu}
\end{align}
\end{subequations}
where $\hat{\Pi}^{0}$ only contains light degrees of freedom and
$\Pihatfullbare$ receives virtual contributions from the heavy quarks.
From Eqs.~\eqref{eq:decgluFull} we then get
\begin{align}
  \zeta_{\Ahat}^0 = \frac{1+\Pihateffbare}{1+\Pihatfullbare} \, .
  \label{eq:decglu2}
\end{align}
Since the l.\,h.\,s\ does not depend on the kinematics of the process
it is possible to consider the special case $p=0$.  The
renormalization of the decoupling constant is done in the usual way by
means of the gluon field renormalization constants in the effective
and the full theory:
$\zeta_{\Ahat}=\bar{Z}^{\phantom{f}}_{\Ahat}/\Zfull_{\Ahat}\,\zeta_{\Ahat}^0$.

The same method also applies to the decoupling of the
$\epsilon$-scalar field where, however, according to the discussion in
Sec.~\ref{sec:epsilonmass} a mass counterterm has to be added in order
to maintain the $\epsilon$-scalar massless. In fact, this counterterm
is even required to ensure that
\begin{align}
  \zeta_{\Atilde}^{0}
  =\frac{1+\Pitildeeffbare}{1+\Pitildefullbare
    +\delta m^2_{\epsilon}}\Bigg|_{p\to 0} 
  \label{eq:decglu3}
\end{align}
is properly defined.

For the calculations in this work we need the decoupling transformations
for $\alpha_{s}$ and $\alpha_{e}$ at the one-loop level which
can be obtained from a matching of the $g q\bar{q}$ and
$\gtilde q\bar{q}$ vertices, in analogy to Eqs.~\eqref{eq:decgluFull}
\begin{equation}
\zeta^0_{g_{s}}
=\frac{1}{\zeta^0_{\psi}\sqrt{\zeta^0_{\Ahat}}}\,
  \frac{1+\Gammaghatfullbare}{1+\Gammaghateffbare\phantom{\Big|}}\, ,\qquad\qquad
\zeta^0_{g_{e}}
=\frac{1}{\zeta^0_{\psi}\sqrt{\zeta_{\Atilde}^{0}}}\,
  \frac{1+\Gammagtildefullbare}{1+\Gammagtildeeffbare\phantom{\Big|}}\, .
\end{equation}
Since $\zeta_{\psi}^0$, $(\Gammaghatfullbare-\Gammaghateffbare)$, and
$(\Gammagtildefullbare-\Gammagtildeeffbare)$ are of
$\mathcal{O}(\alpha^2)$, the (bare) one-loop decoupling constants for
$\gs$ and $\ge$ are entirely given by $\zeta_{\Ahat}^0$ and
$\zeta_{\Atilde}^0$, respectively.  Using
$(\zeta^0_{g_{s}})^2=\zeta^0_{\alpha_{s}}$ and $\zeta_{\alpha_{s}}=
\bar{Z}^{\phantom{f}}_{\alphas}/\Zfull_{\alphas}\,\zeta_{\alpha_{s}}^0$
and similar for the evanescent coupling we finally obtain
\begin{subequations}
\begin{align}
\zeta_{\alphas}
&=1+\Big(\frac{\alpha_{s}}{4\pi}\Big)\,N_{H}\,
  \frac{2}{3}\,\text{ln}\left(\frac{\mu^{2}}{m^{2}}\right)
  +\mathcal{O}(\alpha^2) \, ,\\
\zeta_{\alphae}
&=1+\Big(\frac{\alpha_{e}}{4\pi}\Big)\,N_{H}\,
  \text{ln}\left(\frac{\mu^{2}}{m^{2}}\right)
  +\mathcal{O}(\alpha^2)
\end{align}
\end{subequations}
for the renormalized decoupling constants of $\alphas$ and $\alphae$.

\subsection{Field and mass renormalization of the heavy quarks}
\label{sec:onshell}

\begin{figure}[t]
\begin{center}
\scalebox{.9}{
\begin{picture}(135,50)(0,0)
\DoubleLine[arrow](-10,0)(20,0){2}
\DoubleLine[arrow](20,0)(100,0){2}
\DoubleLine[arrow](100,0)(130,0){2}
\DashCArc(60,0)(40,0,62){4}
\DashCArc(60,0)(40,118,180){4}
\DoubleArc[arrow](60,36)(18,0,180){2}
\DoubleArc[arrow](60,36)(18,180,360){2}
\Vertex(20,0){2}
\Vertex(100,0){2}
\Vertex(42,36){2}
\Vertex(78,36){2}
\end{picture}
\qquad\qquad
\begin{picture}(135,50)(0,0)
\DoubleLine[arrow](-10,0)(20,0){2}
\DoubleLine[arrow](20,0)(100,0){2}
\DoubleLine[arrow](100,0)(130,0){2}
\DashCArc(60,0)(40,0,180){4}
\Vertex(20,0){2}
\Vertex(100,0){2}
\Text(60,32.5)[b]{\scalebox{2}{\ding{53}}}
\end{picture}
}
\end{center}
\caption{\label{epsiloninsertion}
Sample two-loop contributions to the field renormalization of the heavy quark.
The diagram on the r.\,h.\,s.\ shows the insertion of the mass
counterterm $\delta m_{\epsilon}^2$.}
\end{figure}
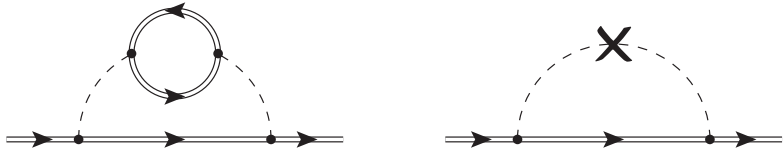

To obtain UV-finite Green functions in the \FDH\ scheme we need to
perform a renormalization of the heavy quark field and mass, where the
corresponding renormalization constants are defined by
\begin{align}
 \psi^0=\sqrt{\bar{Z}_{2,h}}\,\psi,\qquad\quad
 m^0=\bar{Z}_m\, m \, .
\end{align}
Extending the standard \CDR\ procedure for obtaining renormalization constants
in the on-shell (OS) scheme, we write the 1PI self-energy of the heavy quark
in \FDH\ as
\begin{align}
 \bar{\Sigma}(p,m,\Neps)
 = m\,\bar{\Sigma}_{1}(p^2,m,\Neps)
 + (\slashed{p}-m)\,\bar{\Sigma}_{2}(p^2,m,\Neps)\, .
\end{align}
The renormalization constants are then given by
\begin{subequations}
\begin{align}
 \big(\bar{Z}_{2,h}\big)^{-1}
 &=1+2m^2 \frac{\partial}{\partial p^2}\bar{\Sigma}_{1}\big|_{p^2=m^2}
   +\bar{\Sigma}_{2}\big|_{p^2=m^2}\, ,
 \label{eq:Z2OS}
 \\
 \bar{Z}_m
 &=1+\bar{\Sigma}_{1}\big|_{p^2=m^2}\, .
 \label{eq:ZmOS}
\end{align}
\end{subequations}
To obtain their values we calculated the quantities
$\bar{\Sigma}_{1}$ and $\bar{\Sigma}_{2}$ up to the two-loop level,
with sample diagrams shown in Fig.~\ref{epsiloninsertion}.  One point
of major importance is that apart from genuine two-loop diagrams we
have to include contributions originating from UV
(sub)renormalization.  This in particular comprises the mass
counterterm for the $\epsilon$-scalar given in
Eq.~\eqref{eq:epsMassCT}, see the r.\,h.\,s.\ of
Fig.~\ref{epsiloninsertion}.  In terms of the bare couplings we then
get
\begin{align}\label{z2os}
\bar{Z}_{2,h}&=
1+
a_{s}(m^2)\,C_{F}\,\Bigg[
  -\frac{3}{\epsilon}
  -4
  -\epsilon\Big(8+\frac{\pi^2}{4}\Big)
  \Bigg]
+a_e(m^2 )\,C_{F}\,\Neps\,\Bigg[
  -\frac{1}{2\epsilon}
  -\frac{1}{2}
  -\epsilon\Big(\frac{1}{2}+\frac{\pi^2}{24}\Big)
  \Bigg]
\nn\\&\quad
+a_s^2(m^2)\,\Bigg\{
  C_{F}^{2}\,\Bigg[
    \frac{9}{2\epsilon^{2}}
    +\frac{51}{4\epsilon}
    +\frac{433}{8}
    -\frac{49}{4}\pi^{2}
    +16\pi^{2}\ln(2)
    -24\zeta(3)
    \Bigg]
\nn\\&\qquad\qquad\qquad
  +C_{A}C_{F}\,\Bigg[
    -\frac{11}{2\epsilon^2}
    -\frac{101}{4\epsilon}
    -\frac{803}{8}
    +\frac{49}{12}\pi^2
    -8\pi^2\ln(2)
    +12\zeta(3)
\nn\\&\qquad\qquad\qquad\qquad\qquad\
    +\Neps\,\Big(
      \frac{1}{4\epsilon ^2}
      +\frac{11}{8\epsilon}
      +\frac{5}{24}\pi^2
      +\frac{81}{16}
      \Big)
    \Bigg]
\nn\\&\qquad\qquad\qquad
  +C_{F}N_{F}\,\Bigg[
    \frac{1}{\epsilon^2}
    +\frac{9}{2\epsilon}
    +\frac{59}{4}
    +\frac{5}{6}\pi^2
    \Bigg]
  +C_{F}N_{H}\,\Bigg[
    \frac{2}{\epsilon^2}
    +\frac{19}{6\epsilon}
    +\frac{1139}{36}
    -\frac{7}{3}\pi^2
    \Bigg]
  \Bigg\}
\nn\\&\quad
+a_e^2(m^2)\,\Neps\,\Bigg\{
  C_{F}^{2}\,\Bigg[
    \frac{1}{\epsilon^2}
    +\frac{2}{\epsilon}
    +\frac{\pi^2}{2}
    -3
    +\Neps\Big(
      -\frac{1}{8\epsilon^2}
      -\frac{3}{16\epsilon}
      -\frac{13}{48}\pi^2
      +\frac{91}{32}
      \Big)
    \Bigg]
\nn\\&\qquad\qquad\qquad\quad
  +C_{A}C_{F}\,\Bigg[
    \Big(
      -\frac{1}{2\epsilon^2}
      -\frac{1}{\epsilon}
      -\frac{\pi^2}{4}
      +\frac{3}{2}
      \Big)\Big(
      1-\frac{\Neps}{2}
      \Big)\Bigg]
\nn\\&\qquad\qquad\qquad\quad
  +C_{F}N_{F}\,\Bigg[
    \frac{1}{4\epsilon^2}
    +\frac{7}{8\epsilon}
    +\frac{21}{16}
    +\frac{5}{24}\pi^2
    \Bigg]
  +C_{F}N_{H}\,\Bigg[
    \frac{1}{4\epsilon^2}
    +\frac{7}{8\epsilon}
    -\frac{3}{16}
    +\frac{\pi^2}{24}
    \Bigg]
  \Bigg\}
\nn\\&\quad
+a_{s}(m^2)\,a_{e}(m^2)\,\Neps\,\Bigg\{
  C_{F}^{2}\,\Bigg[
    \frac{3}{2\epsilon}
    +\frac{47}{4}
    -\pi^2
    \Bigg]
  +C_{A}C_{F}\,\Bigg[
    -\frac{9}{4\epsilon}
    -\frac{77}{8}
    +\frac{\pi^2}{6}
    \Bigg]
  \Bigg\}
+\mathcal{O}(a^3) \, .
\end{align}
For later purposes it is convenient to introduce a mass counterterm
$\delta m=m-m^{0}=m\,(1-\bar{Z}_{m})$ for the heavy quarks.
Using Eq.~\eqref{eq:ZmOS}, a direct calculation of $\bar{\Sigma}_{1}$
yields
\begin{align}\label{eq:epsMassRen}
\frac{\delta m}{m}&=
a_{s}(m^2)\,C_{F}\,\Bigg[
  \frac{3}{\epsilon}
  +4
  +\epsilon\Big(8+\frac{\pi^2}{4}\Big)
  \Bigg]
+a_e(m^2 )\,C_{F}\,\Neps\,\Bigg[
  \frac{1}{2\epsilon}
  +\frac{1}{2}
  +\epsilon\Big(\frac{1}{2}+\frac{\pi^2}{24}\Big)
  \Bigg]
\nn\\&\quad
+a_s^2(m^2)\,\Bigg\{
  C_{F}^{2}\,\Bigg[
   - \frac{9}{2\epsilon^{2}}
    -\frac{45}{4\epsilon}
    -\frac{199}{8}
    +\frac{17}{4}\pi^{2}
    -8\pi^{2}\ln(2)
    +12\zeta(3)
    \Bigg]
\nn\\&\qquad\qquad\qquad
  +C_{A}C_{F}\,\Bigg[
    \frac{11}{2\epsilon^2}
    +\frac{91}{4\epsilon}
    +\frac{605}{8}
    -\frac{5}{12}\pi^2
    +4\pi^2\ln(2)
    -6\zeta(3)
\nn\\&\qquad\qquad\qquad\qquad\qquad\
    +\Neps\,\Big(
      -\frac{1}{4\epsilon ^2}
      -\frac{9}{8\epsilon}
      -\frac{5}{24}\pi^2
      -\frac{63}{16}
      \Big)
    \Bigg]
\nn\\&\qquad\qquad\qquad
  +C_{F}N_{F}\,\Bigg[
    -\frac{1}{\epsilon^2}
    -\frac{7}{2\epsilon}
    -\frac{45}{4}
    -\frac{5}{6}\pi^2
    \Bigg]
  +C_{F}N_{H}\,\Bigg[
   - \frac{1}{\epsilon^2}
    -\frac{7}{2\epsilon}
    -\frac{69}{4}
    +\frac{7}{6}\pi^2
    \Bigg]
  \Bigg\}
\nn\\&\quad
+a_e^2(m^2)\,\Neps\,\Bigg\{
  C_{F}^{2}\,\Bigg[
    -\frac{1}{\epsilon^2}
    -\frac{3}{\epsilon}
    +\frac{\pi^2}{6}
    -6
    +\Neps\Big(
      \frac{1}{8\epsilon^2}
      +\frac{13}{16\epsilon}
      -\frac{11}{48}\pi^2
      +\frac{75}{32}
      \Big)
    \Bigg]
\nn\\&\qquad\qquad\qquad\quad
  +C_{A}C_{F}\,\Bigg[
    \Big(
      \frac{1}{2\epsilon^2}
      +\frac{3}{2\epsilon}
      -\frac{\pi^2}{12}
      +3
      \Big)\Big(
      1-\frac{\Neps}{2}
      \Big)\Bigg]
\nn\\&\qquad\qquad\qquad\quad
  +C_{F}N_{F}\,\Bigg[
    -\frac{1}{4\epsilon^2}
    -\frac{5}{8\epsilon}
    -\frac{11}{16}
    -\frac{5}{24}\pi^2
    \Bigg]
\nn\\&\qquad\qquad\qquad\quad
  +C_{F}N_{H}\,\Bigg[
    -\frac{1}{4\epsilon^2}
    -\frac{5}{8\epsilon}
    -\frac{3}{16}
    -\frac{\pi^2}{24}
    \Bigg]
  \Bigg\}
\nn\\&\quad
+a_{s}(m^2)\,a_{e}(m^2)\,\Neps\,\Bigg\{
  C_{F}^{2}\,\Bigg[
    \frac{3}{2\epsilon}
    +\frac{23}{4}
    -\pi^2
    \Bigg]
  +C_{A}C_{F}\,\Bigg[
    \frac{3}{4\epsilon}
    +\frac{11}{8}
    +\frac{\pi^2}{2}
    \Bigg]
  \Bigg\}
+\mathcal{O}(a^3) \, .
\end{align}
up to the two-loop level. The pure $\alpha_s$ terms for $\Neps=0$
correspond to the \CDR\ result.

\subsection{Field renormalization of the light quarks}

\begin{figure}
\begin{center}
\scalebox{.9}{
\begin{picture}(135,50)(0,0)
\Line[arrow](-10,0)(20,0)
\Line[arrow](20,0)(100,0)
\Line[arrow](100,0)(130,0)
\GlueArc(60,0)(40,0,62){4}{5}
\GlueArc(60,0)(40,118,180){4}{5}
\DoubleArc[arrow](60,40)(18,0,180){2}
\DoubleArc[arrow](60,40)(18,180,360){2}
\Vertex(20,0){2}
\Vertex(100,0){2}
\Vertex(42,37){2}
\Vertex(78,37){2}
\end{picture}
\qquad\qquad
\begin{picture}(135,50)(0,0)
\Line[arrow](-10,0)(20,0)
\Line[arrow](20,0)(100,0)
\Line[arrow](100,0)(130,0)
\DashCArc(60,0)(40,0,62){4}
\DashCArc(60,0)(40,118,180){4}
\DoubleArc[arrow](60,36)(18,0,180){2}
\DoubleArc[arrow](60,36)(18,180,360){2}
\Vertex(20,0){2}
\Vertex(100,0){2}
\Vertex(42,36){2}
\Vertex(78,36){2}
\end{picture}
}
\end{center}
\caption{\label{lightZ2}
Two-loop contributions to the field renormalization of the light quark.}
\end{figure}
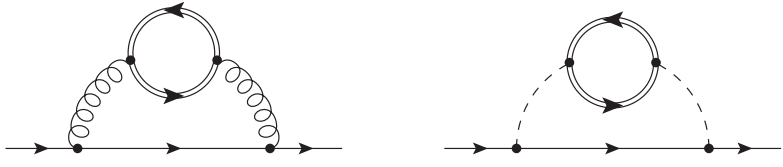

In analogy to the previous section we determine the field renormalization of
the light quark fields where the corresponding renormalization constant is
in the following denoted by $\bar{Z}_{2,l}$.
As in the case of heavy quarks, $\bar{Z}_{2,l}$ receives contributions from
heavy quark loops, see Fig.~\ref{lightZ2}.
However, there is no one-loop contribution since in dimensional regularization
all corresponding loop integrals are scaleless. This also implies that up to the
two-loop level there is no contribution from the $\epsilon$-scalar mass
counterterm. The explicit calculation then yields for the field renormalization
of the light quark in the \FDH\ scheme
\begin{equation}
\bar{Z}_{2,l}=1+
C_{F}N_{H}\Bigg[
  a_s^2(m^2)\Big(
    \frac{1}{2\epsilon }
    -\frac{5}{12}
    \Big)
  +a^2_e(m^2)\,\Neps\,\Big(
    -\frac{1}{4\epsilon^2}
    +\frac{3}{8\epsilon}
    -\frac{13}{16}
    -\frac{\pi^2}{24}
    \Big)
\Bigg]
+\mathcal{O}(a^3) \, .
\end{equation}
As for the mass counterterm, the pure $\alpha_s$ terms are of course
not new.

\section{IR anomalous dimensions in the massive case}
\label{sec:ad}

The aim of this section is to provide all so far unknown IR anomalous
dimensions present in the general IR factorization
formula~\eqref{eq:G2parmass}, i.\,e.\ $\bar{\gamma}_Q$ and
$\bar{\gamma}_{\text{cusp}}(\beta)$.  As in the massless
case~\cite{Broggio:2015dga}, for this we use the SCET framework.

\subsection{Scheme dependence of the heavy-to-light soft function and
$\gamma_{Q}$ }
\label{sec:gQ}
In Ref.~\cite{Gao:2012ja}, it has been shown that the top quark decay
factorizes into regions where only soft radiation and (or) radiation
collinear to the massless partons are present. More precisely, the
factorization consists of a hard function whose renormalization
group equation (RGE) depends on the heavy-quark anomalous dimension, a
quark jet function, and a soft function.  In \CDR, the jet and soft
functions have been calculated up to the two-loop level in
Refs.~\cite{Becher:2006qw} and \cite{Becher:2005pd}, respectively.  In
\FDH, so far only the jet function is known \cite{Broggio:2015dga}.

The general relation between the corresponding IR anomalous dimensions
is given by
\begin{align}\label{gamQ}
\gamma_{Q}^{\mRS}=\gamma_{S}^{\mRS}+\gamma_{J}^{\mRS}-\gamma_{q}^{\mRS}\, ,
\end{align}
where $\gamma_{S}^{\mRS}$ and $\gamma_{J}^{\mRS}$ are the (\RS-dependent)
anomalous dimensions of the soft and
jet function. Eq.~(\ref{gamQ}) is a direct consequence of the fact
that the RGE of the factorization formula does not depend on the
factorization scale.  The values of
$\bar{\gamma}_{J}=\gamma_{J}^{\text{\FDH/\DRED}}$ and $\bar{\gamma}_{q} =
\gamma_{q}^{\text{\FDH/\DRED}}$ have been calculated in
Ref.~\cite{Broggio:2015dga} up to the two-loop level.  In order to
obtain $\bar{\gamma}_Q = \gamma_{Q}^{\text{\FDH/\DRED}}$ we therefore have
to compute $\bar{\gamma}_S = \gamma_{S}^{\text{\FDH/\DRED}}$.

Extending the approach of Ref.~\cite{Becher:2005pd}, we define the
scheme-dependent (bare) soft function as
\begin{align}
   S_{\text{bare}}^{\mRS}\Big( \ln\frac{\Omega}{\mu},\mu \Big)
   \defeq \int_0^{\Omega} d\omega\,
   \langle b_v|\,\bar h_v\,\delta(\omega+in\cdot D)\,h_v\,|b_v\rangle \,,
\end{align}
where $h_v$ are effective quark fields in
HQET~(see e.\,g.\ Ref.~\cite{Neubert:1993mb}),
$b_v$ are on-shell $b$-quark states with velocity $v$,
and $n$ is a light-like 4-vector with $n\cdot v=1$ and $n^2=0$.
The normalization is fixed by $\langle b_v|\,\bar h_v\,h_v\,|b_v\rangle=1$.

For explicit calculations it is useful to express the soft function as a contour
integral
\begin{align}\label{eq:softdisp}
   S_{\text{bare}}^{\mRS}\Big( \ln\frac{\Omega}{\mu},\mu \Big)
   = \frac{1}{2\pi i} \oint\limits_{|\omega|=\Omega} d\omega\,
   \langle b_v|\,\bar h_v\,\frac{1}{\omega+in\cdot D+i0}\,h_v\,|b_v\rangle
   =\frac{1}{2\pi i} \oint\limits_{|\omega|
   =\Omega} d\omega\,\mathcal{S}_{\text{bare}}^{\mRS}\Big(\omega\Big)
\end{align}
and to work in Laplace space
\begin{align}\label{soft:Lap}
s_\text{bare}^{\mRS}(\Omega)
\defeq \int_0^\infty d \omega\, 
\exp\left(-\frac{\omega}{\Omega\,e^{\gamma_E}}\right)
\frac{1}{\pi}\text{Im}\Big{[}\mathcal{S}_{\text{bare}}^{\mRS}(\omega)\Big{]}\, .
\end{align}
Since $h_v$ and $b_v$ are Heisenberg fields, the usual perturbative expansion
results in loop diagrams contributing to the heavy quark propagator.
As in the massless case, the scheme dependence is related to the
UV singularities of such diagrams.

\begin{figure}[t]
\begin{center}
\scalebox{.9}{
\begin{picture}(135,50)(-7,0)
\DoubleLine[arrow](-10,0)(20,0){2}
\DoubleLine[arrow](20,0)(100,0){2}
\DoubleLine[arrow](100,0)(130,0){2}
\GlueArc(60,0)(40,0,62){4}{5}
\GlueArc(60,0)(40,118,180){4}{5}
\DashCArc(60,40)(18,0,360){4}
\Vertex(20,0){2}
\Vertex(100,0){2}
\Vertex(42,37){2}
\Vertex(78,37){2}
\Text(20,-7.5)[b]{\scalebox{2}{\ding{53}}}
\end{picture}
\qquad\
\begin{picture}(135,50)(-7,0)
\DoubleLine[arrow](-10,0)(20,0){2}
\DoubleLine[arrow](20,0)(100,0){2}
\DoubleLine[arrow](100,0)(130,0){2}
\GlueArc(60,0)(40,0,62){4}{5}
\GlueArc(60,0)(40,118,180){4}{5}
\DashCArc(60,40)(18,0,360){4}
\Vertex(20,0){2}
\Vertex(100,0){2}
\Vertex(42,37){2}
\Vertex(78,37){2}
\Text(100,-7.5)[b]{\scalebox{2}{\ding{53}}}
\end{picture}
\qquad\
\begin{picture}(135,50)(-7,0)
\DoubleLine[arrow](-10,0)(20,0){2}
\DoubleLine[arrow](20,0)(60,0){2}
\DoubleLine[arrow](60,0)(100,0){2}
\DoubleLine[arrow](100,0)(130,0){2}
\GlueArc(60,0)(40,0,62){4}{5}
\GlueArc(60,0)(40,118,180){4}{5}
\DashCArc(60,40)(18,0,360){4}
\Vertex(20,0){2}
\Vertex(100,0){2}
\Vertex(42,37){2}
\Vertex(78,37){2}
\Text(60,-7.5)[b]{\scalebox{2}{\ding{53}}}
\end{picture}
}
\end{center}
\caption{\label{fig:massivesoft} Evanescent two-loop contributions to
  the heavy-to-light soft anomalous dimension in the \FDH\ scheme. The
  crosses denote the insertion of the operator $(\omega+in\cdot
  D+i0)^{-1}$.}
\end{figure}
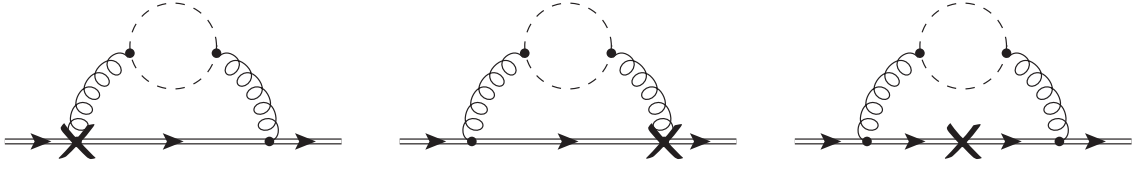

At the one-loop level there are no evanescent contributions since the
$\epsilon$-scalar does not couple to heavy quark lines, see also
Ref.~\cite{Broggio:2015dga}.  There are exactly three diagrams that
induce a scheme dependence of the soft function at the two-loop
level. They are shown in Fig.~\ref{fig:massivesoft}.  For the
explicit computation we generated the diagrams with
QGRAF~\cite{Nogueira:1991ex} and applied a tensor reduction of the
integrals with Reduze~2~\cite{vonManteuffel:2012np}, where the master
integrals needed in \FDH\ are identical to the ones of
\CDR\ given in Ref.~\cite{Becher:2005pd}.

In \FDH\ we then get up to the two-loop level
\begin{align}\label{sbare}
\bar{s}_\text{bare}(\Omega)
   =
   1 + a_{s}(\Omega^2)\, C_F &\Bigg[
    \!-\! \frac{2}{\epsilon^2}
    \!+\! \frac{2}{\epsilon}
    \!-\! \frac{5}{6}\pi^2
   + \epsilon\Big(
     \frac{5}{6}\pi^2
     - \frac{14}{3}\,\zeta_3
     \Big)
   \!-\!\epsilon^2\Big(
     \frac{193}{720}\pi^4
     -\frac{14}{3}\,\zeta_3
     \Big)
    +{\cal O}(\epsilon^3)
    \Bigg]
\nonumber\\*
  +\, a^2_{s}(\Omega^2)
   \, C_F&
   \Bigg[ C_F \bar{K}_F(\epsilon) + C_A \bar{K}_A(\epsilon)
     + \frac{1}{2} N_F \bar{K}_f(\epsilon) 
    \Bigg]+{\cal O}(a^3)\,,
\end{align}
with
\begin{subequations}
\label{eq:sbareCoeff}
\begin{align}
\bar{K}_{F}(\epsilon)&=
  \frac{2}{\epsilon ^4}
  -\frac{4}{\epsilon ^3}
  +\frac{2+\frac{5 \pi ^2}{3}}{\epsilon ^2}
  +\frac{-\frac{10}{3}\pi^2+\frac{28}{3}\zeta(3)}{\epsilon }
  +\frac{5}{3}\pi^2-\frac{56}{3}\zeta(3)+\frac{53}{60}\pi^4\, , \\
\bar{K}_{A}(\epsilon)&=
  -\frac{11}{6 \epsilon ^3}
  +\frac{-\frac{1}{18}+\frac{\pi^2}{6}}{\epsilon ^2}
  +\frac{-\frac{55}{27}-\frac{37}{12}\pi^2+9\zeta(3)}{\epsilon }
  -\frac{326}{81}-\frac{41}{12}\pi^2-\frac{437}{9}\zeta(3)+\frac{107}{180}\pi^4
\nn\\*
  &\quad\,+N_{\epsilon}\Big{(} \frac{1}{12 \epsilon ^3}+
  \frac{1}{18 \epsilon ^2}
  +\frac{\frac{1}{27}+\frac{\pi ^2}{8}}{\epsilon }
  +\frac{2}{81}+\frac{\pi^2}{12}+\frac{25}{18}\zeta(3)\Big)\, ,\\
\bar{K}_{f}(\epsilon)&=
  \frac{2}{3 \epsilon ^3}
  -\frac{2}{9 \epsilon ^2}
  +\frac{-\frac{4}{27}+\pi^2}{\epsilon }
  -\frac{8}{81}-\frac{\pi^2}{3}+\frac{100}{9}\zeta(3) \,.
\end{align}
\end{subequations}
Taking the limit $\Neps\to 0$ in Eq.~\eqref{sbare} we obtain
the \CDR\ result which is in agreement with the one given in
Ref.~\cite{Becher:2005pd}.

As for the quark and gluon jet functions~\cite{Broggio:2015dga},
all divergences of the soft function can be removed multiplicatively
by means of a $Z$ factor
\begin{equation}\label{eq:rensoft}
s_{\text{sub}}^{\mRS}(\Omega,\mu)= 
  Z_{S}^{\mRS}(\Omega,\mu)\,
  s_{\text{bare}}^{\mRS}(\Omega)\, .
\end{equation}
To relate $Z_{S}^{\mRS}(\Omega,\mu)$ with $\gamma_{S}^{\mRS}$ we compare the RGE
of the soft function,
\begin{align}
\label{rge:softZ}
\frac{d}{d\ln\mu}\, s_{\text{sub}}^{\mRS}(\Omega,\mu)=
  \Bigg[
  \Big(\frac{d}{d\ln\mu}\,Z^{\mRS}_{S}(\Omega,\mu)\Big)\,
  \Big(Z^{\mRS}_{S}(\Omega,\mu)\Big)^{-1}
  \Bigg]\,  
  s_{\text{sub}}^{\mRS}(\Omega,\mu)\, \, ,
\end{align} 
with the RGE written in terms of $\gamma_{S}^{\mRS}$,
\begin{align} 
\label{rge:softG}
\frac{d}{d \ln \mu}s_{\text{sub}}^{\mRS}(\Omega,\mu)&=
\Bigg[
  C_F\,\gamma_{\text{cusp}}^{\mRS}\, L_\Omega 
 - 2\gamma_{S}^{\mRS}
 \Bigg]\,
s_{\text{sub}}^{\mRS}(\Omega,\mu) \, ,
\end{align}
where $L_\Omega= \ln(\Omega^{2}/\mu^2)$ and the cusp anomalous dimension is
known from the massless
case~\cite{Gnendiger:2014nxa, Broggio:2015ata,Broggio:2015dga}.
In \FDH, the factor $\bar{Z}_{S}$ is given by
\begin{align}
\ln\bar{Z}_{S}
&=\Big(\frac{\alpha_{s}}{4\pi}\Big)\Bigg[
\frac{C_F\,\bar{\gamma}^{\text{cusp}}_{10}}{2\epsilon^{2}}
-\frac{1}{\epsilon}\Bigg(\frac{C_F\,\bar{\gamma}^{\text{cusp}}_{10}}{2}\,
  L_\Omega-\bar{\gamma}_{10}^{S}\Bigg)\Bigg]
\nonumber\\
&\quad+\Big{(}\frac{\alpha_{s}}{4\pi}\Big{)}^{2}
\Bigg[\,
-\frac{3\,C_F\,\bar{\gamma}^{\text{cusp}}_{10}\,\bar\beta^s_{20}}{8\epsilon^{3}}
+\frac{\bar\beta^s_{20}}{2\,\epsilon^{2}}
  \Bigg(\frac{C_F\,\bar{\gamma}^{\text{cusp}}_{10}}{2}\,L_\Omega
-\bar{\gamma}_{10}^{S}\Bigg)
+\frac{C_F\,\bar{\gamma}^{\text{cusp}}_{20}}{8\,\epsilon^{2}}
\nonumber\\
&\qquad\qquad\quad
-\frac{1}{2\,\epsilon}
\Bigg(\frac{C_F\,\bar{\gamma}^{\text{cusp}}_{20}}{2}\,
  L_\Omega-\bar{\gamma}_{20}^{S}\Bigg)\Bigg]
+{\cal O}(\alpha^3)
\label{eq:logZS}
\end{align}
and the coefficients of the $\beta$ function 
can be found e.\,g.\ in Ref.~\cite{Broggio:2015dga}.
Imposing minimal subtraction with $\Neps$ as an independent quantity
we can read off the soft anomalous dimension
\begin{align}
\bar{\gamma}_{S}
=\ &
\Big{(}\frac{\alpha_{s}}{4\pi}\Big{)}
  \big( -2 C_{F}\big)
\nn\\*
&+\Big{(}\frac{\alpha_{s}}{4\pi}\Big{)}^{2}\Bigg\{
  C_{A}C_{F}\Bigg[
    \frac{110}{27}
    +\frac{\pi^2}{18}
    -18 \zeta(3)
    -\Neps\Big(\frac{2}{27}-\frac{\pi^2}{36}\Big)\Bigg]
  +C_{F}N_{F}\Bigg[
    \frac{4}{27}
    +\frac{\pi^2}{9}\Bigg] \Bigg\}
\nn\\*
    &+\mathcal{O}(\alpha^3) \, ,
\end{align}
which is scheme independent at the one-loop level.
Apart from $\bar{\gamma}_{S}$ it is also possible to extract the already known
values of the cusp anomalous dimension as well as the $\beta$ functions in the
\FDH\ scheme, 
which provides a strong consistency check on the applied procedure.
Using the obtained results together with Eq.~\eqref{gamQ} we then find
\begin{align}\label{anomres}
   \bar{\gamma}_{Q}
   =\ &
   \Big(\frac{\alpha_{s}}{4\pi}\Big)
     \big(-2C_{F}\big)\nn\\*
   &+\Big(\frac{\alpha_{s}}{4\pi}\Big{)}^{2}\Bigg\{
     C_{A} C_{F}\Bigg[
       -\frac{98}{9}
       +\frac{2}{3}\pi ^2
       -4\zeta(3)
       +\frac{8}{9}\Neps\Bigg]
     +C_{F} N_{F}\,\frac{20}{9}\Bigg\}
     +\mathcal{O}(\alpha^3)
\end{align}
for the IR anomalous dimension of the heavy quarks in the \FDH\ scheme.
Like $\bar{\gamma}_S$, at NLO it does not depend on $\Neps$ and is therefore
scheme independent, as already found in Ref.~\cite{Catani:2000ef}.
However, at NNLO it receives \RS-dependent contributions.

Eq.~\eqref{anomres} is the main result of this section. However, for the sake of 
completeness we give the result of the finite and scheme-independent soft function
by setting $\Neps=2\epsilon$ and taking the subsequent limit $\epsilon\to 0$
\begin{align}
s_{\text{fin}}(\Omega,\mu) &=
\lim_{\Neps,\epsilon\to\,0} s_{\text{sub}}^{\mRS}(\Omega,\mu) 
=  1+
\Big(\frac{\alpha_{s}}{4\pi}\Big) \Bigg[
    -C_F\gamma_{10}^{\text{cusp}}\frac{L_\Omega^{2}}{4}
   +\gamma_{10}^{S}L_\Omega+c_{1}^{S}  \Bigg]
\nonumber\\
&\quad +\Big{(}\frac{\alpha_{s}}{4\pi}\Big{)}^{2} \Bigg[
  C_F^{2} (\gamma_{10}^{\text{cusp}})^{2}\frac{L_\Omega^{4}}{32}
  +\Big(
    2\gamma_{10}^{S}\big(\gamma_{10}^{S}-\beta_{20}^{s}\big)
    -C_F(\gamma_{20}^{\text{cusp}}+\gamma_{10}^{\text{cusp}}c_{1}^{S})
    \Big) \frac{L_\Omega^{2}}{4}\phantom{\Bigg|}
\nonumber\\*& \qquad\qquad\quad\ \
    +\Big(\beta^s_{20}-3\gamma_{10}^{S}\Big)\,
          C_F\gamma_{10}^{\text{cusp}}\frac{L_\Omega^{3}}{12}
    +\Big(
      c_{1}^{S}\big(\gamma_{10}^{S}-\beta_{20}^{s}\big)
      +\gamma_{20}^{S}
      \Big)L_\Omega+c_{2}^{S}\Bigg]\, ,
\end{align} 
with
\begin{subequations}
\begin{align}
c_{1}^{S}&=C_{F}\Big{(} -\frac{5\pi^{2}}{6}\Big{)}\, ,\\
c_{2}^{S}&=C^2_{F}\Big{(}  \frac{25 \pi ^4}{72}   \Big{)}
   +C_{F}C_{A}\Big(
     -\frac{326}{81}
     -\frac{233 \pi ^2}{36}
     -\frac{283 \zeta (3)}{9}
     +\frac{107 \pi ^4}{180}\Big)\nonumber\\
&\quad+C_{F}N_F\Big(
  -\frac{4}{81}+\frac{7}{18}\pi^2+\frac{22}{9}\zeta(3)\Big) \, .
\end{align}
\end{subequations}
This result is in agreement with the one given in Ref.~\cite{Becher:2005pd}.

\subsection{Determination of $\bar{ \gamma}_{\rm cusp}(\beta)$
}
\label{sec:gC}

\begin{figure}[t]
\begin{center}
\scalebox{.9}{
\begin{picture}(135,65)(0,10)
\DoubleLine[arrow](0,65)(50,65){2}
\DoubleLine[arrow](50,65)(100,65){2}
\DoubleLine(0,75)(0,55){3}
\Gluon( 50, 0)( 50, 65){5}{6}
\Vertex( 50, 65){2}
\Text(90,73)[l]{\scalebox{1.11}{$p_I$}}
\LongArrow( 60, 45)( 60, 25)
\Text( 65, 35)[l]{\scalebox{1.11}{$k\rightarrow 0$}}
\end{picture}
\quad
\begin{picture}(135,65)(0,10)
\DoubleLine[arrow](0,65)(50,65){2}
\DoubleLine[arrow](50,65)(100,65){2}
\DoubleLine(0,75)(0,55){3}
\DashLine( 50, 65)( 50, 0){4}
\Vertex( 50, 65){2}
%
\LongArrow( 60, 45)( 60, 25)
\Text( 65, 35)[l]{\scalebox{1.11}{$k\rightarrow 0$}}
\end{picture}
}
\end{center}
\caption{\label{fig:eikonalFeynmanRules} Coupling of a gluon (left)
  and an $\epsilon$-scalar (right) to a heavy quark propagator. In the
  eikonal approximation the latter vanishes.  }
\end{figure}
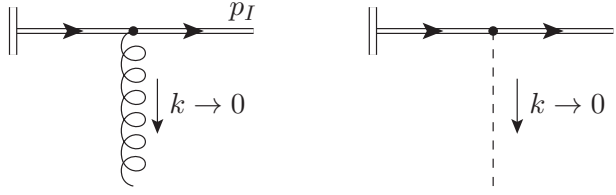

The velocity-dependent cusp anomalous dimensions can be extracted from the
heavy-to-heavy soft anomalous dimension $\Gamma_{hh}$ for the pair production of
massive quarks. Using \CDR, $\Gamma_{hh}$ has been calculated in
Ref.~\cite{Kidonakis:2009ev} in the framework of the eikonal approximation.
This method can also be used to derive the respective quantity in \FDH.

In general, the eikonal approximation is suited for describing the emission
of soft gluons from partons in a hard scattering process,
see the l.\,h.\,s.\ of Fig.~\ref{fig:eikonalFeynmanRules}.
For a vanishing gluon momentum, the Feynman rule for the coupling of a gluon to
a massive quark propagator can be reduced to
\begin{subequations}
\begin{align}
 \bar{u}(p_I)(-i\gs T^a)\,\gammahat^\mu
 \left[i \frac{\slashed{p}_I+\slashed{k}+m_{I}^{\phantom{|}}}
   {(p_I+k)^2-m_{I}^{2}}\right]
 \ \rightarrow & \
 \bar{u}(p_I)\,\gs T^a\,\gammahat^\mu
 \left[ \frac{\slashed{p}_I+m_{I}^{\phantom{|}}}{2\, p_I\cdot k}\right]\\
 \ = & \
 \bar{u}(p_I)\,\gs T^a\,
 \left[ (p_I)_{\nu}\frac{\{\gammahat^\mu,\gammahat^\nu\}}{2\, p_I\cdot k}\right]\\
 \ = & \
 \bar{u}(p_I)\,\gs T^a\,
 \left[\frac{v_I^\mu}{v_I\cdot k}\right]\, ,
 \label{eq:CouplingEikonal}
\end{align}
\end{subequations}
where in the second line the Dirac equation
$\bar{u}(p_I)(\slashed{p}_I-m_I^{\phantom{|}})=0$ has been
used.
Since the Feynman rule~\eqref{eq:CouplingEikonal} does not contain a
Dirac matrix anymore, the evaluation of loop contributions is much simpler compared
to ordinary QCD.

Extending this to the case of an $\epsilon$-scalar we get
\begin{align}
 \bar{u}(p_I)(-i\ge T^a)\,\gammatilde^\mu
 \left[i \frac{\slashed{p}_I+\slashed{k}+m_{I}^{\phantom{|}}}
   {(p_I+k)^2-m_{I}^{2}}\right]
 &\ \rightarrow \
 \bar{u}(p_I)\,\ge T^a\,
 \left[ (p_I)_{\nu}\frac{\{\gammatilde^\mu,\gammahat^\nu\}}{2\, p_I\cdot k}\right]
 =0\, .
 \label{eq:evCouplingEikonal}
\end{align}
Due to the vanishing anticommutator, a direct coupling of $\epsilon$-scalars to
massive quark propagators does not exist in the eikonal approximation.

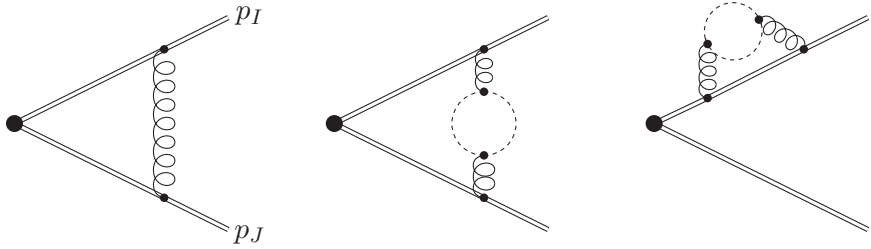
\begin{figure}[t]
\begin{center}
\scalebox{.8}{
\begin{picture}(135,90)(20,0)
\DoubleLine(30,45)(100,80){2}
\DoubleLine(100,80)(130,95){2}
\DoubleLine(30,45)(100,10){2}
\DoubleLine(100,10)(130,-5){2}
\Gluon(100,10)(100,80){5}{7}
\Text(140,100)[t]{\scalebox{1.25}{$p_I$}}
\Text(140, -3)[t]{\scalebox{1.25}{$p_J$}}
\Vertex(30,45){4}
\Vertex(100,80){2}
\Vertex(100,10){2}
\end{picture}
\quad
\begin{picture}(135,90)(20,0)
\DoubleLine(30,45)(100,80){2}
\DoubleLine(100,80)(130,95){2}
\DoubleLine(30,45)(100,10){2}
\DoubleLine(100,10)(130,-5){2}
\Gluon(100,60)(100,80){4}{2}
\Gluon(100,10)(100,30){5}{2}
\DashCArc(100,45)(15,-90, 90){2}
\DashCArc(100,45)(15, 90,270){2}
\Vertex(30,45){4}
\Vertex(100,60){2}
\Vertex(100,30){2}
\Vertex(100,80){2}
\Vertex(100,10){2}
\end{picture}
\quad
\begin{picture}(135,90)(20,0)
\DoubleLine(30,45)(55,57){2}
\DoubleLine(55,57)(100,80){2}
\DoubleLine(100,80)(130,95){2}
\DoubleLine(30,45)(130,-5){2}
\DashCArc(67,88.5)(13.5, 15,195){2}
\DashCArc(67,88.5)(13.5,195,375){2}
\Gluon(55,57)(55,82.5){4}{3}
\Gluon(79,94)(100,80){4}{3}
\Vertex(30,45){4}
\Vertex(55,57){2}
\Vertex(100,80){2}
\Vertex(55,82.5){2}
\Vertex(79,94){2}
\end{picture}
}
\end{center}
\caption{\label{fig:eikonalDiagrams} One- and two-loop contributions
  to the heavy-to-heavy soft anomalous dimension in the eikonal
  approximation. Since there is no direct coupling of
  $\epsilon$-scalars to massive quark propagators there is no
  evanescent contribution at the one-loop level.  }
\end{figure}
Following the approach of Ref.~\cite{Kidonakis:2009ev}, the soft anomalous
dimension for heavy-quark pair production can be obtained from the UV poles of
corresponding eikonal diagrams with one- and two-loop examples shown in
Fig.~\ref{fig:eikonalDiagrams}.
Since there is no direct coupling of $\epsilon$-scalars to massive
quarks, the soft anomalous dimension is scheme independent at the one-loop level.
At the two-loop level, however, closed $\epsilon$-scalar loops yield evanescent
contributions $\propto\alphas \Neps$.

In the following, the scalar product of the two outgoing velocity vectors is fixed
by $v_I\cdot v_J \defeq -\,\text{cosh}\,\beta_{IJ}$ with $\beta_{IJ}$ given
in Eq.~\eqref{eq:betaIJ}, and the indices of $\beta$ are suppressed:
$\beta_{IJ}\eqdef\beta$.
Generalizing Eq.~(14) of Ref.~\cite{Becher:2009kw} to the case of \FDH,
the result of the soft anomalous dimension can then be written as
\begin{align}
 \bar\Gamma_{hh}(v_I,v_J)
 = C_F\,\bar{ \gamma}_{\rm cusp}(\beta)+2\,\bar{\gamma}_Q \, .
\end{align}


Using Eq.~\eqref{anomres}, it is now possible to extract the velocity-dependent
cusp anomalous dimension in \FDH\ which in terms of the renormalized
couplings reads
\begin{align}
  \bar{ \gamma}_{\rm cusp}(\beta)
   =\ & \bar{\gamma}_{\rm cusp}\,\beta\coth\beta
     +8\,C_A \left( \frac{\alphas}{4\pi} \right)^2\Bigg\{
     \beta^2 + \frac{\pi^2}{6} + \zeta_3
   \nn\\*
   &+\coth\beta^{\phantom{2}} \Bigg{[} 
     \mbox{Li}_2(e^{-2\beta}) - 2\,\beta\,\ln(1-e^{-2\beta}) 
     -\frac{\pi^2}{6}\,(1+\beta) - \beta^2 - \frac{\beta^3}{3}\Bigg{]}
   \nn\\
   &+\coth^2\beta \Bigg{[} \mbox{Li}_3(e^{-2\beta}) 
     + \beta\,\mbox{Li}_2(e^{-2\beta})
     -\zeta_3 + \frac{\pi^2}{6}\,\beta + \frac{\beta^3}{3} \Bigg{]}
   \Bigg\}+\mathcal{O}(\alpha^3) \, .
\label{eq:gCuspBeta}
\end{align}
Since the terms in the curly brackets do not depend on $\Neps$,
the scheme dependence of $\bar{ \gamma}_{\rm cusp}(\beta)$
is entirely governed by the scheme dependence of the cusp anomalous dimension 
in the massless case, i.\,e.\ $\bar{ \gamma}_{\rm cusp}$.

\section{Guideline for FDH calculations and checks of the results
}
\label{sec:examples}

In order to check the obtained results for the scheme dependence of IR
divergences in massive QCD we compute the heavy and the heavy-to-light quark
form factor in \FDH\ up to the two-loop level. Apart from a pure check this
section is also intended to provide a guideline how practical calculations in
the \FDH\ scheme can actually be done. For the two-loop calculations we
therefore use the following approach:
\begin{itemize}
 \item At the one-loop level we distinguish the $\epsilon$-scalar from the
   $D$-dimensional gluon since the related couplings $\alphas$ and $\alphae$
   renormalize differently.
 \item At the two-loop level we use a (quasi) $4$-dimensional Lorentz algebra
   for the evaluation of genuine two-loop diagrams and do not distinguish the
   $\epsilon$-scalar from the $D$-dimensional gluon.
 \item After having applied the UV renormalization we set equal the couplings
   $\alphas$ and $\alphae$ in contributions from one-loop counterterm diagrams.
 \item Throughout the calculations we identify $\Neps=2\epsilon$.
\end{itemize}
Using this setup it turns out that practical calculations in the \FDH\ scheme
are not significantly more complicated than the respective ones in \CDR.

\subsection{Heavy quark form factor}

\begin{figure}[t]
\begin{center}
\scalebox{.9}{
\begin{picture}(135,90)(0,10)
\Photon(0,45)(30,45){3}{3}
\DoubleLine[arrow](30,45)(100,80){2}
\DoubleLine[arrow](100,80)(120,90){2}
\DoubleLine[arrow](120,0)(100,10){2}
\DoubleLine[arrow](100,10)(30,45){2}
\Gluon(100,10)(100,80){4}{9}
\Vertex(30,45){2}
\Vertex(100,80){2}
\Vertex(100,10){2}
%
\LongArrow(105,90)(115,95)
\Text(110,102)[c]{\scalebox{1.1}{$p_1$}}
\LongArrow(110,12)(120,7)
\Text(115,20)[c]{\scalebox{1.1}{$p_2$}}
\end{picture}
\qquad
\begin{picture}(135,90)(0,10)
\Photon(0,45)(30,45){3}{3}
\DoubleLine[arrow](30,45)(100,80){2}
\DoubleLine[arrow](100,80)(120,90){2}
\DoubleLine[arrow](120,0)(100,10){2}
\DoubleLine[arrow](100,10)(30,45){2}
\DashLine(100,10)(100,80){4}
\Vertex(30,45){2}
\Vertex(100,80){2}
\Vertex(100,10){2}
\end{picture}
}
\end{center}
\caption{\label{fig1}
One-loop diagrams contributing to the heavy-quark form factor in \FDH.}
\end{figure}
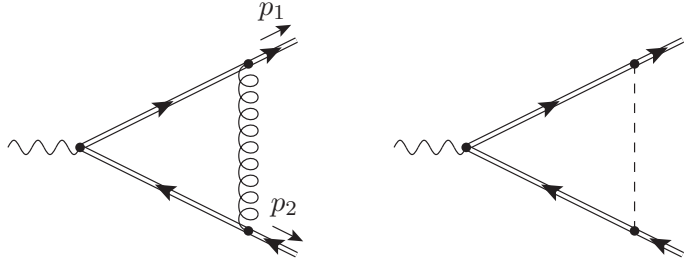

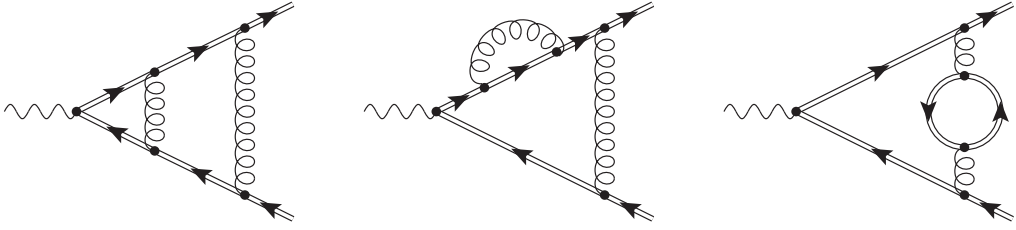
\begin{figure}[t]
 \scalebox{.9}{
\begin{picture}(135,90)(0,0)
\Photon(0,45)(30,45){3}{3}
\DoubleLine[arrow](30,45)(62.5,61.5){2}
\DoubleLine[arrow](62.5,61.5)(100,80){2}
\DoubleLine[arrow](100,80)(120,90){2}
\DoubleLine[arrow](120,0)(100,10){2}
\DoubleLine[arrow](100,10)(62.5,28.5){2}
\DoubleLine[arrow](62.5,28.5)(30,45){2}
\Gluon(62.5,28.5)(62.5,61.5){4}{4}
\Gluon(100,10)(100,80){4}{9}
\Vertex(30,45){2}
\Vertex(62.5,61.5){2}
\Vertex(62.5,28.5){2}
\Vertex(100,80){2}
\Vertex(100,10){2}
\end{picture}
\quad
\begin{picture}(135,90)(0,0)
\Photon(0,45)(30,45){3}{3}
\DoubleLine[arrow](30,45)(50,55){2}
\DoubleLine[arrow](50,55)(80,70){2}
\DoubleLine[arrow](80,70)(100,80){2}
\DoubleLine[arrow](100,80)(120,90){2}
\DoubleLine[arrow](120,0)(100,10){2}
\DoubleLine[arrow](100,10)(30,45){2}
\Gluon(100,10)(100,80){4}{9}
\GlueArc(65,62.5)(17, 20, 200){4}{6}
\Vertex(30,45){2}
\Vertex(50,55){2}
\Vertex(80,70){2}
\Vertex(100,80){2}
\Vertex(100,10){2}
\end{picture}
\quad
\begin{picture}(135,90)(0,0)
\Photon(0,45)(30,45){3}{3}
\DoubleLine[arrow](30,45)(100,80){2}
\DoubleLine[arrow](100,80)(120,90){2}
\DoubleLine[arrow](120,0)(100,10){2}
\DoubleLine[arrow](100,10)(30,45){2}
\Gluon(100,60)(100,80){4}{2}
\Gluon(100,10)(100,30){4}{2}
\DoubleArc[arrow](100,45)(15,90,270){2}
\DoubleArc[arrow](100,45)(15,270,90){2}
\Vertex(30,45){2}
\Vertex(100,80){2}
\Vertex(100,60){2}
\Vertex(100,30){2}
\Vertex(100,10){2}
\end{picture}
}
\caption{\label{fig2}
Sample two-loop diagrams contributing to the heavy-quark form factor in \FDH.
All gluons belong to the quasi 4-dimensional space $Q4S$.}
\end{figure}

In \CDR, the heavy quark form factor has been calculated up to NNLO in
Ref.~\cite{Bernreuther:2004ih}. 
In \FDH, the Green function for the interaction of a
virtual photon and two massive quarks can be written as
\begin{subequations}
\begin{align}
\hspace{-5mm}
  \bar{V}^{\mu}_{c_{1}c_{2}}(p_1,p_2) & =
    \bar{u}_{c_{1}}(p_1)\,
    \bar{\Gamma}^{\mu}_{c_{1}c_{2}}(p_1,p_2)\,
    v_{c_{2}}(p_2) \, ,
\end{align}
with
\begin{align}
  \bar{\Gamma}^{\mu}_{c_{1}c_{2}}(p_1,p_2)
  & = - i \,v_Q \, \delta_{c_{1}c_{2}} \Biggl[
    \bar{F}_{1}(x) \, \hat{\gamma}^{\mu} 
  + \frac{1}{2m} \bar{F}_{2}(x) \, i \, \hat{\sigma}^{\mu \nu} q_{\nu}
  \Biggr] \,.
\label{b0002} 
\end{align}
\end{subequations}
Here and in the following, $p_1$ and $p_2$ denote the (outgoing) momenta of the two
external quarks with $p_{1}^{2}=p_{2}^{2}=m^2$ and $s = (p_{1}+p_{2})^2/m^2$.
In general, the $\gamma$ matrices appearing in Eq.~\eqref{b0002} are
scheme-dependent. However, since we are only interested in the structure of
$\bar{F}_1$ their dimensionality can be chosen arbitrarily.
Here and in the following we therefore use $D$-dimensional $\gamma$ matrices
in the Lorentz decomposition.

The IR anomalous dimensions can be obtained from the heavy-quark form factor,
$\bar{F}_1$, which can be extracted from Eq.~\eqref{b0002} by using an appropriate
projection operator. For the proper definition of the projection and other
details we refer to Ref.~\cite{Bernreuther:2004ih}.
In the \FDH\ scheme, only two diagrams contribute to the form factor at the
one-loop level, see Fig.~\ref{fig1}. Using 1-dimensional harmonic polylogarithms
\cite{Remiddi:1999ew,Gehrmann:2001pz} of the variable
\be
x = \frac{\sqrt{-s+4} - \sqrt{-s} }{\sqrt{-s+4} + \sqrt{-s} } 
\qquad ( 0 \leq x \leq 1 )
\label{xvar}
\ee
and notation \eqref{eq:asdef} for the couplings, we represent the
bare one-loop coefficients of the form factor as%
\footnote{Note that $\bar{F_1}$ denotes the (all-order) form factor in \FDH\ whereas
its perturbative coefficients are written using a calligraphic form,
$\big[\bar{F_1}\big]_{mn}=\bar{\mathcal{F}}_{mn}$.}
\bea
\hspace{-5mm}
\bar{F}_{1}(x) & = & 
   1 + a_{s}(m^{2})\,\bar{\mathcal{F}}_{10}(x)
     + a_{e}(m^{2})\,\bar{\mathcal{F}}_{01}(x)
     + {\mathcal O}(a^{2})\, ,
     \label{b00020} 
\eea
with 
\begin{subequations}
\begin{align}
\bar{\mathcal{F}}_{10}(x) & =   
2\,C_{F} \, \Biggl\{\frac{1}{\epsilon}  \,\Biggl[
   \frac{1}{2}
   +H(0;x)\,\frac{x^2+1}{x^2-1}
   \Biggr]
   +\frac{1}{2} H(0;x)\,\frac{x+1}{x-1}
\nn\\&\qquad\qquad
   -\Bigg(
     \frac{\pi^2}{6}
     -H(0;x)
     -H(0,0;x)
     +2 H(\!-1,0;x)
     \Bigg)\,\frac{x^2+1}{x^2-1}\phantom{\Bigg|}
\nn\\&\qquad\qquad
+ \epsilon \,\Bigg[\,
  \frac{\pi^2}{24}
  -\Bigg(
     \frac{\pi^2}{12}
     - \frac{H(0,\!0;x)}{2} 
     + H(\!-1,\!0;x)
     \Bigg)\frac{x+1}{x-1}
   -\Bigg(
     \frac{\pi^2}{6}
     -\big(4\! - \! \frac{\pi^2}{12}\big) H(0;x)
\nn\\&\qquad\qquad\qquad
   + 2\,\zeta(3)
   - \frac{\pi^2}{3}\,H(\!-1;x) \! 
   - H(0,\!0;x)
   + 2 H(\!-1,\!0;x)
   - H(0,\!0,\!0;x)\phantom{\Bigg|}
\nn\\*&\qquad\qquad\qquad
   + 2 H(\!-1,\!0,\!0;x)
   + 2 H(0,\!-1,\!0;x) \! 
   - 4 H(\!-1,\!-1,\!0;x)
   \Bigg)\frac{x^2+1}{x^2-1}\!\Bigg]
+\mathcal{O}(\epsilon^2)\Bigg\}, 
\label{1loopF1}
\\
\bar{\mathcal{F}}_{01}(x) & = 
    \, C_{F} \,\Bigg\{
     1+\epsilon\,\Big[
       1+\frac{1-x}{1+x}H(0;x)\Big]
+\,\mathcal{O}(\epsilon^2)\Bigg\} \, .
\label{1loopF2}
\end{align}
\end{subequations}
To obtain the result at the two-loop level we evaluate the Feynman
diagrams (see Fig.~\ref{fig2}) using a quasi $4$-dimensional Lorentz
algebra.  This in particular means that the absolute number of
diagrams and master integrals~\cite{Bonciani:2003te, Bonciani:2003hc} we have
to evaluate is exactly the same as in \CDR. In line with that we do
not have to introduce evanescent couplings like $\alphae$ in the
computation of the genuine two-loop diagrams.

In the following we give the difference between the UV renormalized form factors
in \FDH\ and \CDR\ at the two-loop level.
For the renormalization of the couplings, the quark mass, and the fields we use
Eqs.~\eqref{eq:asdef}, \eqref{eq:epsMassRen}, and \eqref{z2os}, respectively,
and set $\alphas=\alphae$ after renormalization.
Because of the appearing $\epsilon$-scalar propagator in the right diagram of
Fig.~\ref{fig1} we also have to add the mass counterterm~\eqref{eq:epsMassCT}
of the $\epsilon$-scalar. Combining all results we finally get
\begin{align}
\bar{F}_1(x)-F_{1}(x)&=
\Big(\frac{\alpha_s}{4\pi}\Big)^{2}\Bigg\{
  C_A C_F \Bigg[
  \frac{1}{3\epsilon}-\frac{8}{9}\Bigg]
  \Big(-1+\frac{x^2+1}{x^2-1}H(0;x)\Big{)}
  \nn\\
&\qquad\qquad\quad
  +C_F^{2}\Bigg[
    \frac{2 \left(x^2-1\right) H(0;x)
    -4 \left(x^2+1\right) H(0,0;x)}{(x+1)^2}
    \Bigg]
  +\mathcal{O}(\epsilon^1)
  \Bigg\}
  \nn\\
&\qquad
  +\mathcal{O}(\alphas^3) \, .
\end{align}

This difference can be expressed in terms of the IR anomalous dimensions and
$\beta$ functions through Eqs.~\eqref{eq:G2parmass} and \eqref{eq:ZmassiveFDH},
in a similar way as shown in Ref.~\cite{Broggio:2015dga} for the case of massless
partons:
\begin{align}\label{massdiff}
\bar{F_1}(x)-F_1(x)&=
  \Big{(}\frac{\alpha_s}{4\pi}\Big{)}^{2}\Bigg{\{ }
    -\frac{1}{\epsilon^2}\,C_F\,
    \Big(\bar{\beta}_{20}^s-\beta_{20}^s\Big)
    \Big(-1+\frac{x^2+1}{x^2-1} H(0;x)\Big)
\nn\\ &\qquad\qquad\quad
   +\frac{1}{4\epsilon} \Bigg[
     C_F \Big(
       \bar{\gamma}^{\text{cusp}}_{20}(\beta)
       -\gamma^{\text{cusp}}_{20}(\beta)
       -8\, \mathcal{F}^{\text{diff}}_1
       \Big)
     +2\,\Big(\bar{\gamma}^{Q}_{20}-\gamma^{Q}_{20}\Big)
\nn\\ &\qquad\qquad\qquad\qquad    
  +8\,C_F\,\mathcal{F}^{\text{diff}}_{1}\,\frac{x^2+1}{x^2-1}\,H(0;x)
   \Bigg]+\mathcal{O}(\epsilon^1)
\Bigg{\} }
+\mathcal{O}(\alphas^3)\, ,
\end{align}
where $\mathcal{F}^{\text{diff}}_{1}=
\bar{\mathcal{F}}_{10}^{\text{ren}}
+\bar{\mathcal{F}}_{01}^{\text{ren}} -\mathcal{F}_{1}^{\text{ren}}$ is
the difference of the UV renormalized one-loop coefficients,
i.\,e.\ including a field renormalization of the heavy quarks.  The
fact that the scheme dependence of the IR divergences related to the
heavy-quark form factor can be predicted with the results from
Secs.~\ref{sec:schemes} and \ref{sec:ad} constitutes a strong
consistency check of the results obtained so far.

\subsection{Heavy-to-light form factor}

\begin{figure}[t]
 \scalebox{.9}{
\begin{picture}(135,90)(0,0)
\Photon(0,45)(30,45){3}{3}
\Line[arrow](30,45)(62.5,61.5)
\Line[arrow](62.5,61.5)(100,80)
\Line[arrow](100,80)(120,90)
\DoubleLine[arrow](120,0)(100,10){2}
\DoubleLine[arrow](100,10)(62.5,28.5){2}
\DoubleLine[arrow](62.5,28.5)(30,45){2}
\Gluon(62.5,28.5)(62.5,61.5){4}{4}
\Gluon(100,10)(100,80){4}{9}
\Vertex(30,45){2}
\Vertex(62.5,61.5){2}
\Vertex(62.5,28.5){2}
\Vertex(100,80){2}
\Vertex(100,10){2}
\end{picture}
\quad
\begin{picture}(135,90)(0,0)
\Photon(0,45)(30,45){3}{3}
\Line[arrow](30,45)(50,55)
\Line[arrow](50,55)(80,70)
\Line[arrow](80,70)(100,80)
\Line[arrow](100,80)(120,90)
\DoubleLine[arrow](120,0)(100,10){2}
\DoubleLine[arrow](100,10)(30,45){2}
\Gluon(100,10)(100,80){4}{9}
\GlueArc(65,62.5)(17, 20, 200){4}{6}
\Vertex(30,45){2}
\Vertex(50,55){2}
\Vertex(80,70){2}
\Vertex(100,80){2}
\Vertex(100,10){2}
\end{picture}
\quad
\begin{picture}(135,90)(0,0)
\Photon(0,45)(30,45){3}{3}
\Line(30,45)(120,90)
\Line[arrow](30,45)(100,80)
\Line[arrow](100,80)(120,90)
\DoubleLine[arrow](120,0)(100,10){2}
\DoubleLine[arrow](100,10)(30,45){2}
\Gluon(100,60)(100,80){4}{2}
\Gluon(100,10)(100,30){4}{2}
\DoubleArc[arrow](100,45)(15,90,270){2}
\DoubleArc[arrow](100,45)(15,270,90){2}
\Vertex(30,45){2}
\Vertex(100,80){2}
\Vertex(100,60){2}
\Vertex(100,30){2}
\Vertex(100,10){2}
\end{picture}
}
\caption{\label{lifig1}
Sample two-loop diagrams contributing to the heavy-to-light form factor in \FDH.}
\end{figure}
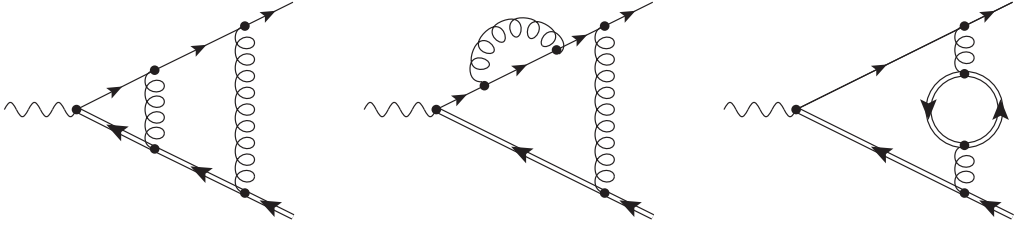

The \CDR\ result for the decay process $b\to u\,W^{*}\to u\,l\,\bar{\nu}_{l}$
has been computed at NNLO in Refs.~\cite{Bonciani:2008wf,
  Asatrian:2008uk, Beneke:2008ei, Bell:2008ws}. 
Applying the procedure of the previous section we here extend the calculation
to the case of \FDH, with sample two-loop diagrams shown in Fig.~\ref{lifig1}. 

In \FDH, the tensor structure of the heavy-to-light form factor can be
written as
\bea 
\bar{\Gamma}^{\mu}(p_1,p_2) &=&
\bar{F}_1(q^2)\,\gammahat^{\mu}
+ \frac{1}{2 \mb} \bar{F}_2(q^2)\,\hat{\sigma}^{\mu \nu}\,q_{\nu}
+ \frac{i}{2 \mb} \bar{F}_3(q^2)\,q^{\mu}
+ \bar{G}_1(q^2)\,\hat{\gamma}^{\mu}\,\gamma_{5}
\nn\\* & &
+ \frac{i}{2 \mb} \bar{G}_2(q^2)\,\gamma_{5}\,q^{\mu}
+ \frac{i}{2 \mb} \bar{G}_3(q^2)\,\gamma_{5}\,(p_1^\mu-p_2^\mu) \, ,
\label{Vmu}
\eea
with $q=p_1+p_2$.
Again, we are interested in the form factor $\bar{F}_{1}$ which can be extracted
by means of a projection operator. Accordingly, the matrix $\hat{\gamma}^{\mu}$
is treated in $D$ dimensions.

We compute the bare diagrams up to NNLO and perform the UV renormalization exactly
in the same way as described in the previous section, taking into account that here
only one leg is massive. Again we have to add a counterterm to subtract
the $\epsilon$-scalar mass shift.
Using Eq.~\eqref{b00020} for the perturbative expansion of the form factor
and expressing the result in terms of the dimensionless quantity
\begin{align}
 y\defeq\frac{q^2}{\mb^2} \, ,
\end{align}
we get for the bare one-loop coefficients
\begin{subequations}
\begin{align}
\hspace*{-5mm}  \bar{\mathcal{F}}_{10}(y) & =
-\,C_{F} \Bigg[
  \frac{1}{\epsilon ^2}
  +\frac{1+2\,H(1;y)}{\epsilon }
  +4
  +\frac{\pi^2}{12}
  +3\,H(1;y)
  +2\,H(0,1;y)
  +4\,H(1,1;y)
\nn\\*&\qquad\qquad
  +\epsilon\, \Bigg(
    8
    +\frac{\pi^2}{12}
    -\frac{\zeta(3)}{3}
    +\Big(8+\frac{\pi^2}{6}\Big)\,H(1;y)
    +3\,H(0,1;y)
    +6\,H(1,1;y)
\nn\\&\qquad\qquad\qquad\
    +8\,H(1,1,1;y)
    +4\,H(-1,0,-1;-y)
    +4\,H(0,-1,-1;-y)
    \phantom{\Bigg|}
\nn\\*&\qquad\qquad\qquad\
    +2\,H(0,0,1;y)
  \Bigg)\Bigg]
  + \, {\mathcal O} \left( \epsilon^2 \right)  \, ,
\label{he1loopF1} \\
\hspace*{-5mm}  \bar{\mathcal{F}}_{01}(y) & =
C_{F} \Bigg[
  1
  +\epsilon\Big(1+H(1;y)\Big)
\Bigg]
+ \, {\mathcal O} \left( \epsilon^2 \right) \, .
\label{he1loopF2}
\end{align}
\end{subequations}
%
%
As in the previous section we give the difference between the
UV renormalized form factors in \FDH\ and \CDR\ up to the two-loop level:
\begin{align}
\bar{F}_{1}(y)&-F_{1}(y)=
\Big{(}\frac{\alpha_s}{4\pi}\Big{)}\frac{C_F}{2}\nn\\
&+\Big{(}\frac{\alpha_s}{4\pi}\Big{)}^2\Bigg\{
C_A C_F\, \Bigg[
  -\frac{1}{4 \epsilon^2}
  +\frac{-\frac{25}{36}-\frac{1}{3} H(1;y)-\frac{L}{6}}{\epsilon }
  +\frac{965}{216}+\frac{\pi^2}{24}+\frac{8}{9} H(1;y)+\frac{4}{9}L\Bigg]
\nn\\&\qquad\qquad\quad
-C_F^2\,\Bigg[
  \frac{1}{2 \epsilon ^2}
  +\frac{\frac{9}{4}+2 H(1;y)+L}{\epsilon }
  +\frac{49}{8}
  +\frac{\pi^2}{4}
  +\big(6+4 L\big) H(1;y)
\nn\\&\qquad\qquad\qquad\qquad\
  +8 H(1,1;y)
  +2 H(0,1;y)
  +\frac{7}{2}L
  +L^2\Bigg]
\nn\\&\qquad\qquad\quad
+C_F N_F\,\Bigg[\frac{1}{4 \epsilon }-\frac{3}{8}\Bigg]
-C_F N_H\,\frac{ L}{2}
+\mathcal{O}(\alphas^3) \, ,
\label{eq:HtoL}
\end{align}
where $L$ is defined as $L=\ln\Big{(}\frac{\mu^{2}}{m^{2}}\Big{)}$.
In terms of the IR anomalous dimensions, the $\beta$ functions, and the
factor $\bar{\mathbf{Z}}$ defined in Eq.~\eqref{eq:ZmassiveFDH} this difference
is given by
\begin{align}\label{eq:tranhtl}
\bar{F_1}(y) & -F_1(y)=
\Big{(}\frac{\alpha_{s}}{4\pi}\Big{)}
  \frac{\bar{\gamma}^{q}_{01}}{2\epsilon}
\nn\\&
+\Big{(}\frac{\alpha_{s}}{4\pi}\Big{)}^{2}\Bigg{\{ }
  \frac{3}{16\epsilon^{3}}\,C_F\gamma^{\text{cusp}}_{10}
    \Big(\bar{\beta}^{s}_{20}-\beta^{s}_{20}\Big)
  -\frac{1}{16\epsilon^{2}} \Bigg{[}
    \bar{\gamma}^{q}_{01}\Big(
      4(\bar{\beta}^{e}_{11}+\bar{\beta}^{e}_{02})
      +2\bar{\gamma}^{q}_{01}
      -4\gamma^{Q}_{10}
      \Big)
\nn\\&\qquad\qquad\quad\ \
    +\Big(\bar{\beta}^{s}_{20}-\beta^{s}_{20}\Big)\Big(
      4\,\big(\gamma^{Q}_{10}+\gamma^{q}_{10}\big)
      -2\,C_F\gamma^{\text{cusp}}_{10}\big(2\,H(1;y)+L\big)
      \Big)
\nn\\&\qquad\qquad\quad\ \
  +C_F \Big(
    \bar{\gamma}^{\text{cusp}}_{20}
    \!-\gamma^{\text{cusp}}_{20}
    \!-8 \bar{\gamma}^{q}_{01}
    \Big)
  +4\,C_F \gamma^{\text{cusp}}_{10} \mathcal{F}^{\text{diff}}_{1}
   \Bigg]
\nn\\&\qquad\qquad\ \
   +\frac{1}{4\epsilon} \Bigg{[}
     -\frac{1}{2} C_F \big(2\,H(1;y)+L\big)
       \Big(
         \bar{\gamma}^{\text{cusp}}_{20}
         -\gamma^{\text{cusp}}_{20}
         +2\,\gamma^{\text{cusp}}_{10}\,\mathcal{F}^{\text{diff}}_{1}
         \Big)
\nn\\&\qquad\qquad\quad\ \
   +(\bar{\gamma}^{Q}_{20}-\gamma^{Q}_{20})
   +(\bar{\gamma}^{q}_{20}-\gamma^{q}_{20})
   +\bar{\gamma}^{q}_{11}
   +\bar{\gamma}^{q}_{02}
   -2\,N_H\,\bar{\gamma}^{q}_{01}\,L\phantom{\Big|}
\nn\\&\qquad\qquad\quad\ \
   +2\,\mathcal{F}^{\text{diff}}_{1}\,\Big(
     \gamma^{Q}_{10}
     +\gamma^{q}_{10}
     +\bar{\gamma}^{q}_{01}
     \Big)
   +2\,\bar{\gamma}^{q}_{01}\,\mathcal{F}^{\text{fin}}_{1}
  \Bigg{]}
  +\mathcal{O}(\epsilon^1)\Bigg{\} }
+\mathcal{O}(\alphas^3)\, ,
\end{align}
with
\begin{subequations}
\begin{align}
\mathcal{F}^{\text{diff}}_{1}&=
  \bar{\mathcal{F}}_{10}^{\text{ren}}
  +\bar{\mathcal{F}}_{01}^{\text{ren}}
  -\mathcal{F}_{1}^{\text{ren}}  \, ,
\\*
\mathcal{F}^{\text{fin}}_{1} &=
  \lim_{\epsilon\to 0}\bigg{[}
    \bar{\mathcal{F}}_{10}^{\text{ren}}
    +\delta\bar{\textbf{Z}}_{10}
    \bigg{]}=
  \lim_{\epsilon\to 0}\bigg{[}
    \mathcal{F}_{10}^{\text{ren}}
    +\delta\textbf{Z}_{1}
    \bigg{]}\, .
\end{align}
\end{subequations}
The fact that Eq.~\eqref{eq:HtoL} matches with Eq.~\eqref{eq:tranhtl}
constitutes an additional and independent check of our results for the
IR anomalous dimensions.

\section{Conclusions}
\label{sec:conc}

The scheme dependence of massless QCD amplitudes at NNLO had been
discussed in Ref.~\cite{Broggio:2015dga}. In this paper we complete
this study by extending it to amplitudes containing massive quarks.

This requires modifications in the UV and IR sector. For the UV part,
the presence of heavy quarks modifies the renormalization. In
particular, the $\epsilon$-scalar field requires a mass
counterterm. Also, the decoupling of $\alpha_e$ (the coupling of the
$\epsilon$-scalars to the quarks) has to be determined.  Furthermore,
we have computed the additional contributions required in \FDH\ in
the quark mass and the quark wave-function renormalization.

Regarding the IR part, the important result is that the IR structure
of massive QCD amplitudes in \FDH\ (and \DRED) is the same as in
\CDR\ (and \HV). The only change is in the explicit scheme-dependent
expressions of the various anomalous dimensions. In the massive case,
there are two additional anomalous dimensions,  the
velocity-dependent cusp anomalous dimension and the heavy-quark anomalous
dimension. We have computed them in the \FDH\ scheme, using a SCET
approach.  

We have checked our results by computing the heavy-quark and
heavy-to-light form factor in \FDH\ at NNLO. These results differ from
the corresponding expressions in \CDR. After UV renormalization, the
difference can be reproduced by the scheme dependence of the IR
factorization formula. This provides us with a strong consistency
check and establishes \FDH\ as a consistent regularization scheme also
in the massive case, at least to NNLO.

\bigskip

\section*{Acknowledgments} 
It is a pleasure to thank Alessandro Broggio and Dominik
St$\breve{\text o}$ckinger for useful discussions and comments on the
manuscript.  We are grateful to Pierpaolo Mastrolia, Thomas Gehrmann
and Andrea Ferroglia for providing assistance with the master
integrals needed for the computation of the form factors and to Thomas
Becher for clarifications about the computation of the soft
function. A.~Visconti is supported by the Swiss National Science
Foundation (SNF) under contracts 200021-144252 and 200021-163466.

\bigskip


\bibliography{bibliography}{}
\bibliographystyle{JHEP}

\end{fmffile}
\end{document}